\documentclass[%
 reprint,
superscriptaddress,
 amsmath,amssymb,
 aps,
 prf,
 onecolumn,
 longbibliography,
]{revtex4-2}

\usepackage{graphicx}
\usepackage{dcolumn}
\usepackage{bm}

\usepackage{gensymb}
\usepackage[utf8]{inputenc}
\usepackage{hyperref}
\usepackage{color}
\usepackage{ulem}
\usepackage{soul}

\begin{document}

\preprint{APS/123-QED}
\title{Rapid wetting of shear-thinning fluids}
\thanks{A footnote to the article title}%

\author{Susumu Yada}
\affiliation{FLOW Centre, Dept. of Engineering Mechanics, Royal Institute of Technology  (KTH),  100 44 Stockholm, Sweden}%
\author{Kazem Bazesefidpar}
\affiliation{FLOW Centre, Dept. of Engineering Mechanics, Royal Institute of Technology  (KTH),  100 44 Stockholm, Sweden}%
\author{Outi Tammisola}
\affiliation{FLOW Centre, Dept. of Engineering Mechanics, Royal Institute of Technology  (KTH),  100 44 Stockholm, Sweden}%
\author{Gustav Amberg}
\affiliation{FLOW Centre, Dept. of Engineering Mechanics, Royal Institute of Technology  (KTH),  100 44 Stockholm, Sweden}%
\affiliation{S\"odert\"orn University,  SE-141 89 Stockholm, Sweden}%
\author{Shervin Bagheri}
\affiliation{FLOW Centre, Dept. of Engineering Mechanics, Royal Institute of Technology  (KTH),  100 44 Stockholm, Sweden}%

\date{\today}

\begin{abstract}
Using experiments and numerical simulations, we investigate the spontaneous spreading of droplets of aqueous glycerol (Newtonian) and aqueous polymer (shear-thinning) solutions on smooth surfaces. 
We find that in the first millisecond the spreading of the shear-thinning solutions is identical to the spreading of water, regardless of the polymer concentration. In contrast, aqueous glycerol solutions show a different behavior, namely, significantly slower spreading rate than water. 
In the initial rapid spreading phase, the dominating forces that can resist the wetting are inertial forces and contact-line friction. For the glycerol solutions, an increase in glycerol concentration effectively increases the contact-line friction, resulting in increased resistance to wetting.  For the polymeric solutions, however, an increase in  polymer concentration does not modify contact-line friction. 
As a consequence, the energy dissipation at the contact line can not be controlled by varying the amount of additives for shear-thinning fluids. 
%
%
%
%
%
%
The reduction of the spreading rate of shear-thinning fluids on smooth surfaces in the rapid wetting regime can only be achieved by increasing solvent viscosity. Our results have implications for phase-change applications where the control of the rapid spreading rate is central, such as anti-icing and soldering.  
\end{abstract}

\maketitle

\section{Introduction}
The motion of a fluid-fluid interface over a solid surface is a challenging problem since the macroscopic behavior of the system depends on the atomistic details of the surface and the fluids. Compared to Newtonian fluids, non-Newtonian liquids have a more complex molecular structure that may influence the triple-phase contact line physics. For example, self-assembly of particles \citep{LU_ACIS201643} and molecular migration from a high-shear region \citep{Ma2005, HAN2013} are features that modify wetting dynamics. To accurately predict and control complex fluids, it is necessary to understand their wetting and spreading on surfaces. For many processes such as inkjet printing, coating, additive manufacturing \citep{LU_ACIS201643} and deposition processes \citep{Bergeron2000}, the wetting dynamics of non-Newtonian fluids is central.

Experimental observations have revealed that both shear-thinning and elastic effects of non-Newtonian fluids modify the wetting dynamics. The high shear rate near the moving contact line results in a small viscous force of shear-thinning liquids, which reduces the viscous bending of the liquid-vapor interface compared to Newtonian fluids \citep{SeevaratnamPoF2007, XWang_Langmuir_2018}.  On the other hand, it has been reported that the fluid elasticity of a Boger fluid enhances the viscous bending of the liquid-vapor interface near the contact line \citep{WEI_JCIS_2007274}. 
A reduced (enhanced) bending of the interface results in a weaker (stronger) dependence of the apparent contact angle on the contact line speed. Indeed, both an increase and a decrease in contact line speeds have been reported for non-Newtonian fluids. \citet{Wei_JPCM_2009} reported that 0.15 wt~\% Xanthan gum solution spreads significantly faster than Newtonian polydimethylsiloxane solution.  On the other hand, \citet{Rafai2004} reported slower spreading of a shear-thinning droplet than the pure Newtonian solvent. 
These findings are also in qualitative agreement with the numerical investigation by \citet{WangPRE2015} using Giesekus droplets which include elastic effect and shear-thinning effect. The majority of experimental \citep{Rafai2004, RAFAI200558, SeevaratnamPoF2007, MINJCIS2010250, WangLangmuir2007, XWang_Langmuir_2018} and modeling  \citep{WeinderPoF1994, NeogiJCP2001, LiangLangmuir2010} investigations involving wetting of non-Newtonian fluids have focused on the wetting regime where viscous resistance dominates, i.e. a droplet slowly approaching equilibrium or steady meniscus on moving plates.

In contrast, the rapid wetting regime of non-Newtonian liquids, which involves non-equilibrium wetting phenomena at small time scales ($\sim$ milliseconds), has been studied much less. The rapid initial spreading of non-Newtonian droplets is of particular importance in applications where phase change occurs, such as anti-icing and soldering. For example, \citet{deRuiter2017} recently reported that a drop stops spreading on a cold surface when the contact line reaches the critical temperature to freeze, in a time scale of a few milliseconds. In such situations, the rapid wetting alone determines the final shape of the droplet. 

The initial stage of spreading is much faster than the rate predicted by Tanner's law \cite{Tanner1979}, which is based on spreading resistance from viscous forces only. In rapid wetting, liquid inertia and contact-line friction also influence the spreading rate \cite{Do-Quang2015}. 
The resistance to spreading contributed by the contact-line friction is related to energy dissipation at the contact line. It depends both on the surface properties (such as adsorption/desorption) and on the liquid properties, in particular viscosity \citep{BLAKE20061}. We refer the reader to the reviews of \citep{Bonn2009, Snoeijer2013} for further details. 
The contact-line dissipation can effectively be represented by an appropriate finite slip length at the contact line \citep{BocquetFD1999}. A more explicit representation is based on the contact-line friction parameter $\mu_f$. This parameter can be directly measured \citep{Duvivier2011, Vo2018, Hong2013Langmuir, Steen2020} or estimated by parameter fitting numerical simulations to experiments \citep{CarlsonPRE2012, Lee2019}.  
The precise relationship between $\mu_f$ and liquid viscosity is not fully established even for Newtonian liquids. In Molecular Kinetic Theory (MKT), where wetting is described as a thermally-activated process of molecular events, a linear relationship has been suggested between contact-line friction and viscosity \cite{BLAKE20061, Haring}. 
However, studies of rapid wetting based on continuum simulations and experiments have observed a sub-linear relationship between contact-line friction and viscosity \citep{CarlsonPRE2012, Vo2018, Hong2013Langmuir, Steen2020}.

This paper aims to increase our understanding of how viscoelastic effects influence rapid wetting. The
paper is organized as follows. In section II, we describe the experimental configuration and observations. Section
III describes the mathematical model of non-Newtonian fluids and the numerical method based on the phase-field method. The section also presents a parametric study of droplet spreading for different total viscosity, line-friction
parameter and relaxation time of polymers. Section IV combines experiments and numerical results to establish
the relative importance of inertia, viscosity, and contact-line friction in the wetting dynamics. Section V discusses
our interpretation of why the rapid spreading of polymer solutions is independent of polymer concentration. Finally,
concluding remarks are provided in section VI.

\begin{figure*}[t!]
    \centering
        \includegraphics[width=0.40\textwidth]{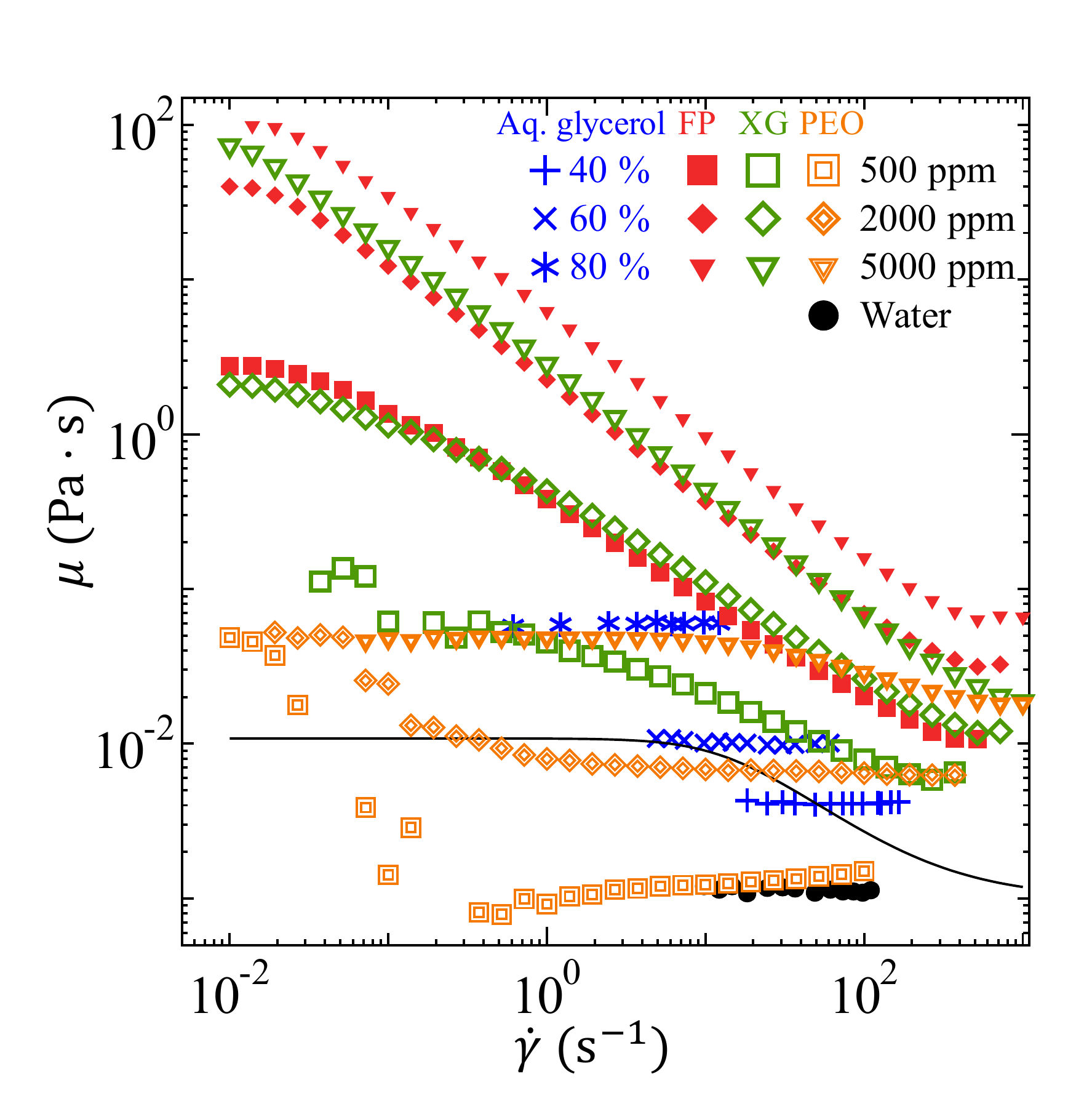}\\
    \caption{Shear viscosity as function of the shear rate $\dot \gamma$. The solid black line represents the Giesekus viscosity based on Eq.~\ref{eq:Viscosity} with $\mu_0 = 0.01$~Pa$\cdot$s (Gi:7, see TABLE \ref{table:simulations}). 
    }
    \label{fig:rheology}
\end{figure*}

\section{Experimental observations}
%
\begin{figure*}[b]
    \centering
    \includegraphics[width=0.8\textwidth]{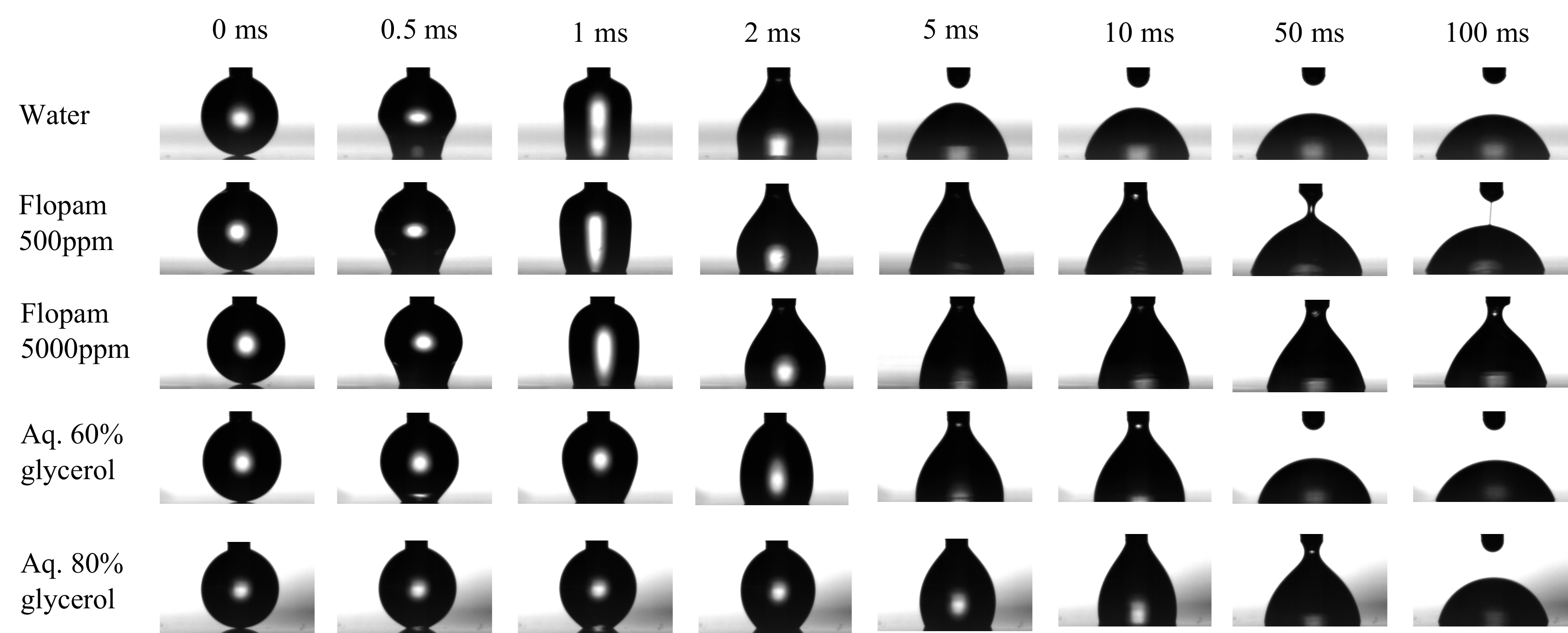}
    \caption{Selected serial snapshots of droplet spreading of water, FLOPAM solutions, and aqueous glycerol solutions. 
    }
    \label{fig:snaps}
\end{figure*}

The fluids in this study are dilute solutions of rigid polymer (Xanthan gum, G1253, Sigma Aldrich), anionic polyacrylamide-based flexible polymer (FLOPAM, AN934SH, SNF), and PEO (molecular weight $4\times 10^6$, 189464, Sigma Aldrich) in deionized water. In addition, aqueous glycerol and deionized water are studied as Newtonian fluids. 
Figure~\ref{fig:rheology} shows the shear viscosity of the investigated liquids as a function of the shear rate. We observe that Xanthan gum (green symbols), FLOPAM (red), and PEO (orange) solutions exhibit shear-thinning viscosity, while water (black) and aqueous glycerol (blue) have constant viscosity over the full range of shear rates. The shear viscosity was measured with Kinexus pro+(NETZSCH). 

\begin{table}
    \centering
    \begin{tabular}{l|cclcc|ll}
    \hline
       & $\sigma$\hspace{6mm}  & $\theta_e$  &\hspace{5mm}$\mu_0$ & $\mu_{\infty}$ & $\mu_{f}$ & $Oh$ & $Oh_f$ 
    \\  & (mN/m)  & ($\deg$)  &\hspace{3mm}(Pa$\cdot$s)  & (Pa$\cdot$s) & (Pa$\cdot$s) & &
    \\ \hline
    Water & 72 & 52  & - & $1.0 \times 10^{-3}$  & 0.12 & $5.3 \times 10^{-3}$ & 0.63 \\ 
    FP 500 ppm  & 72 & 58 & 2.8 & $1.1\times 10^{-2}$  &0.12 & $ 5.8 \times 10^{-2}$& 0.63 \\
     FP 2000 ppm  & 72 & 55 &  $3.9 \times 10^{1}$ & $3.2\times 10^{-2}$ & 0.12& $1.7 \times 10^{-1}$ & 0.63\\
      FP 5000 ppm  & 71 & 58 & $9.9 \times 10^{1}$  & $6.3 \times 10^{-2}$  & 0.12 & $3.3 \times 10^{-1}$ & 0.64\\
       XG 500 ppm  & 70 & 57 & $1.1 \times 10^{-1}$ & $5.8\times 10^{-3}$  &0.12 & $3.1 \times 10^{-2}$ &0.64 \\
     XG 2000 ppm  & 65 & 55 & 2.1 & $1.2\times 10^{-2}$ &0.12& $6.7 \times 10^{-2}$ & 0.67\\
      XG 5000 ppm  & 60 & 58 & $7.3 \times 10^{1}$ & $1.8 \times 10^{-2}$  &0.12 &$1.0 \times 10^{-1}$ &0.69\\
        PEO 500 ppm  & 61 & 49 & $4.4 \times 10^{-2}$ & $1.1 \times 10^{-3}$ &0.12 & $6.3 \times 10^{-3}$ &0.69\\
     PEO 2000 ppm  & 61 & 49 & $5.1 \times 10^{-2}$ & $6.2 \times 10^{-3}$ &0.12& $3.6 \times 10^{-2}$ &0.69\\
      PEO 5000 ppm  & 61 & 49 & $5.3 \times 10^{-2}$  & $1.8 \times 10^{-2}$ &0.12& $1.0 \times 10^{-1}$ &0.69\\
     40\% aq. glycerol  & 65 & 57 & - & $4.0 \times 10^{-3}$  & 0.21& $2.2 \times 10^{-2}$ &1.1\\
      60\% aq. glycerol  & 64 & 55 & - & $1.1\times 10^{-2}$  & 0.33& $6.2 \times 10^{-2}$ & 1.7\\
       80\% aq. glycerol  & 63 & 57 & - & $6.0\times 10^{-2}$  & 0.80 &$3.2 \times 10^{-1}$ &$4.1$\\
    \hline
    \end{tabular}
    \caption{Surface tension $\sigma$, static contact angle $\theta_e$, zero-shear and high-shear viscosity $\mu_0$, $\mu_{\infty}$, the line friction parameter $\mu_f$, the conventional Ohnesorge number based on $\mu$, and the friction Ohnesorge number $Oh_f$. The line friction parameter of water and aqueous glycerol solutions are estimated by matching the spreading curves in the numerical simulations to the rapid spreading experiments.\cite{CarlsonPRE2012, Yada_langmuir_2021} The line friction parameters of polymer solutions are assumed to be identical to water. 
    }
   \label{table:merged}
\end{table}

Table~\ref{table:merged} lists five dimensional quantities of the droplets; the surface tension ($\sigma$), static contact angle ($\theta_e$), zero-rate viscosity ($\mu_0$), the high-shear-rate viscosity ($\mu_{\infty}$) and the contact-line friction ($\mu_f$).
The static contact angle of all liquids is similar, ranging from 49$^{\circ}$ to 58$^{\circ}$. Also, the surface tension stays relatively constant for all solutions, ranging from $60$ to $72$~mN/m.
The surface tension was measured using TD 2 tensiometer (LAUDA).
The zero-rate shear viscosity $\mu_0$ varies significantly for different polymer solutions, whereas the variation of the high-shear rate viscosity $\mu_{\infty}$ is smaller. 

Figure~\ref{fig:snaps} shows a sequence of snapshots from the experiments for different Newtonian and non-Newtonian liquids. 
The droplets with an initial radius of 0.5 $\pm$ 0.02~mm spread on a smooth hydrophilic Off-Stoichiometry-Thiol-Ene (Ostemer 220, Mercene Labs, Sweden) \citep{Carlborg2011} surface. We have used a high-speed camera (speedsense, Dantec Dynamics) with 25,000 frames per second and a spatial resolution of 8~$\mathrm{\mu m}$. The liquid droplet grows from a needle with an outer diameter of 0.31~mm (Hamilton, gauge 30, point style 3) pumped by a syringe pump (Cetoni, neMESYS 1000N) at a flow rate of 0.04 $\mu$L/s. The flow rate is so small that a quasi-static state is assumed before the droplet touches the surface. The droplet starts to spread immediately after it makes contact with the substrate at $t = 0$~ms.
Note that we observe finite apparent contact radius even before the contacts due to the limited spatial resolution.
The first four columns of figure~\ref{fig:snaps} show the spreading in the rapid regime. After this initial phase, the spreading toward equilibrium shapes is much slower. Focusing on the rapid regime, we note that the droplet shape of water is similar to the droplet shapes of the two FLOPAM solutions. The glycerol solutions, on the other hand, spread slower; the solution with 80$\%$ glycerol (bottom row in the figure) has retained much of its spherical shape, while water and polymer solutions have an inverted vase shape.


%

\begin{figure*}[t!]
    \centering
        \includegraphics[width=0.24\textwidth]{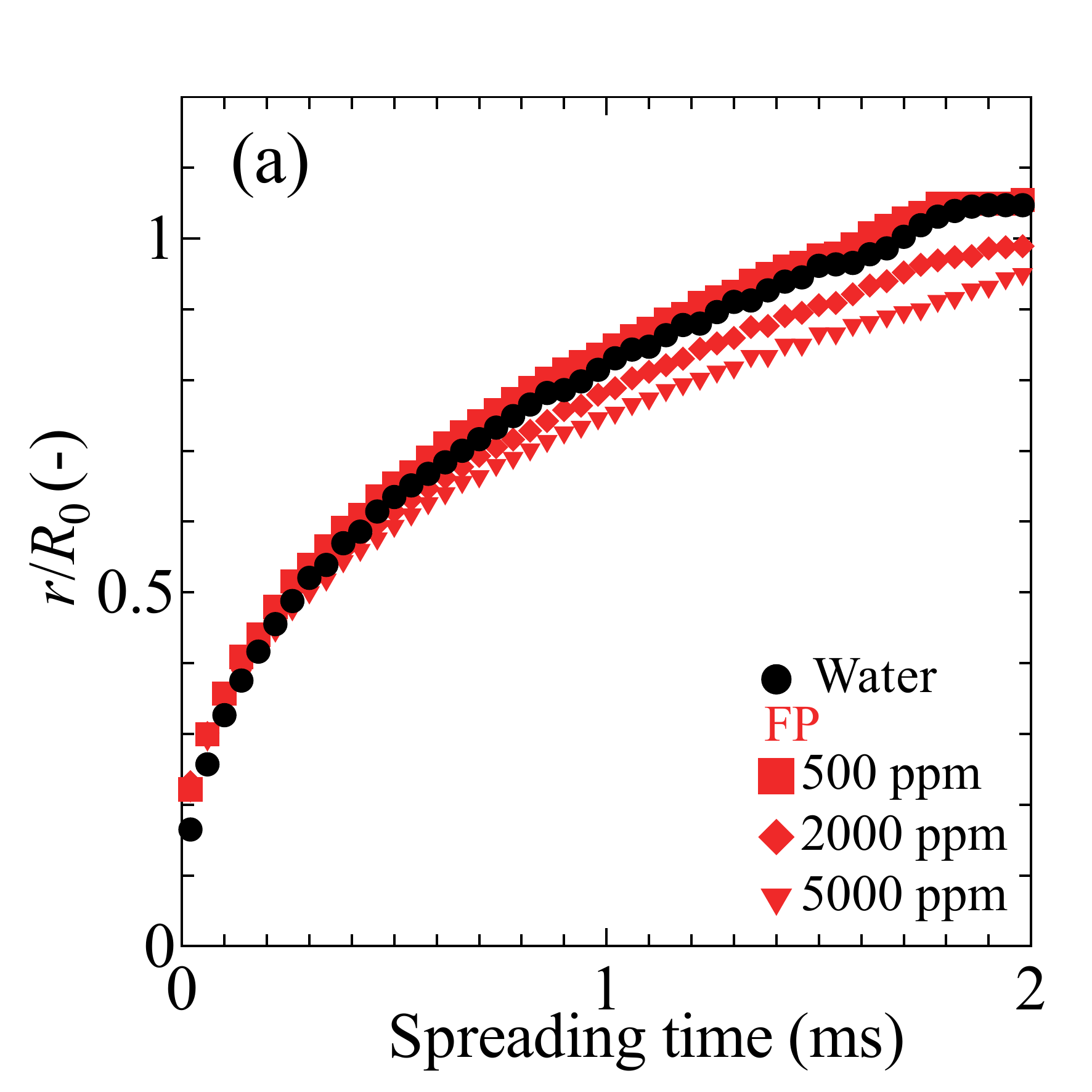}
         \includegraphics[width=0.24\textwidth]{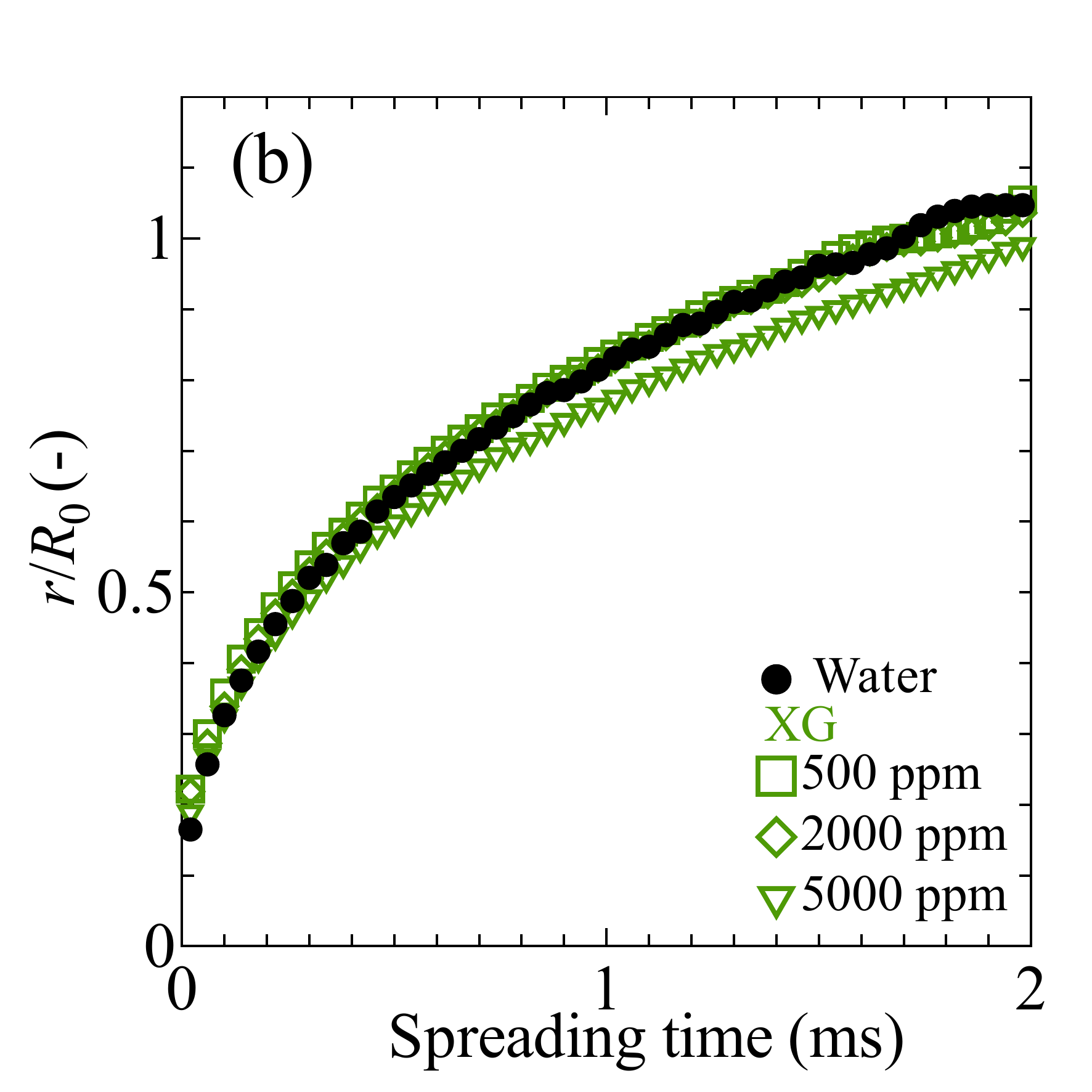}
         \includegraphics[width=0.24\textwidth]{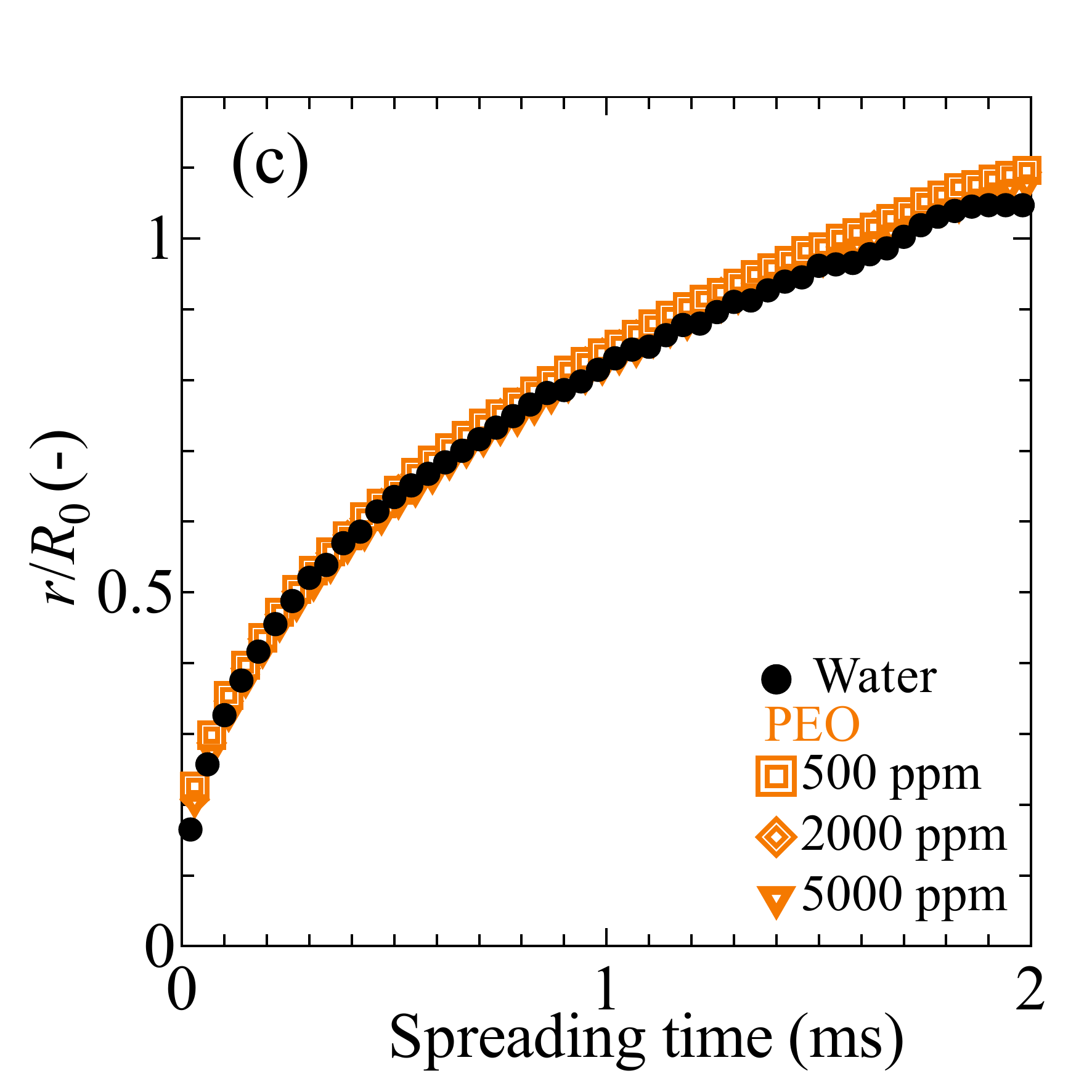}
         \includegraphics[width=0.24\textwidth]{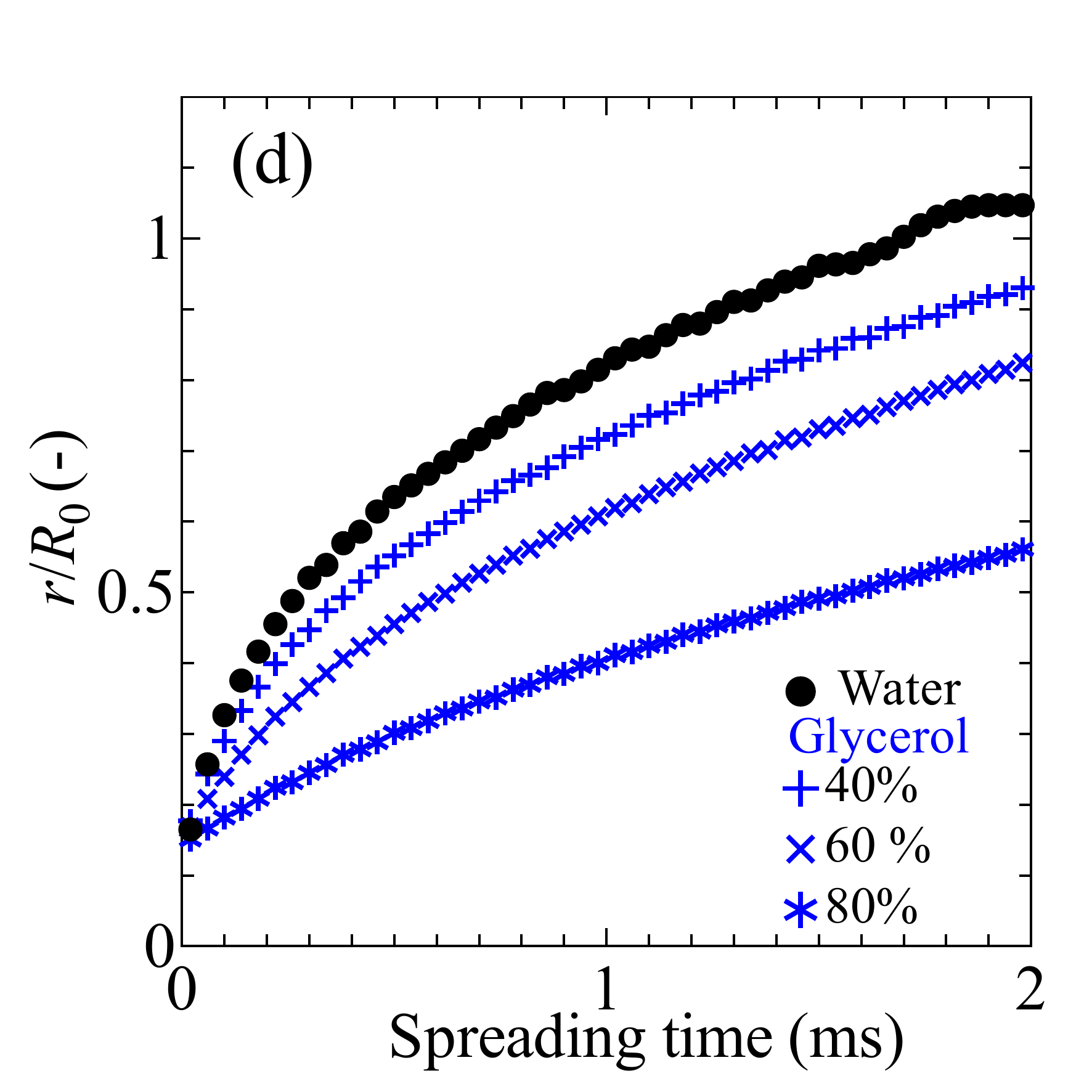}
    \caption{Spreading curves of FLOPAM("FP", red), Xanthan gum("XG", green), polyethylene oxide ("PEO", orange) solutions and aq.~glycerol (blue). The black marks represent water. 
    }
    \label{fig:Data}
\end{figure*}

A more quantitative picture is provided in figure~\ref{fig:Data}. The figure shows the time evolution of the spreading radius with different polymer/glycerol concentrations in the rapid spreading phase. The 500~ppm and 2000~ppm solutions of FLOPAM and Xanthan gum spread similarly to water, while the 5000~ppm solutions spread slightly slower than water (the red and green triangles in Fig.~\ref{fig:Data} a,b). PEO solutions spread very similarly to water, regardless of the concentrations (Fig.~\ref{fig:Data}c). Figure~\ref{fig:Data}(d) shows how glycerol solutions (blue markers) exhibit a reduced spreading rate as the glycerol concentration increases.

\section{Viscoelastic effects and contact-line friction}
The spontaneous spreading of a droplet toward equilibrium is driven by surface tension, and its speed is set by the resistance from inertial forces, viscoelastic stresses, and contact-line friction. The resistive force that dominates depends on the fluid and surface properties and droplet size. To determine the contributions of different sources of resistance, we have performed numerical simulations of water, glycerol, and Giesekus droplets. The latter include both shear-thinning and elastic effects.

\subsection{Diffusive interface model of droplets}
The employed model uses a phase-field variable $\phi$ ranging from 1 in the liquid to -1 in the vapour phase. The variable is governed by the Cahn-Hilliard equation,
\begin{eqnarray}
    &&\frac{\partial  \phi }{\partial t} + \mathbf{u}\cdot \nabla \phi
    = \nabla \cdot \left(M \nabla G( \phi)\right ),
    \label{eq:PF}
\end{eqnarray}
where $M$ is a mobility parameter and $G$ is the chemical potential. The latter is defined by
\begin{equation}
G (\phi) = \frac{3 \sigma \eta}{2\sqrt{2}} (\Psi^{\prime}(\phi)-  \nabla^2 \phi ),
\label{eq:G}
\end{equation}
where $\sigma$ and $\eta$ are surface tension and the diffuse interface thickness, respectively. 
The function $\Psi(\phi)$ in Eq.~\ref{eq:G} represents the standard double-well function, i.e. $\Psi(\phi)= (\phi+1)^{2}(\phi-1)^{2}/\eta^2$.
The derivation of \ref{eq:PF} and \ref{eq:G} can be found in \cite{CarlsonJFM2011, jacqmin_2000}. 
The fluid velocity $\mathbf{u}$ of the two phases is governed by
the incompressible Navier-Stokes equations,
\begin{eqnarray}
 \label{eq:NS}
    &&\rho(\phi) \frac{D \mathbf{u} }{D t} +\mathbf{J} \cdot \nabla \mathbf{u}
    = - \nabla p + \nabla \cdot \left(\frac{(1+\phi)}{2}\mathbf{\tau}\right) + \nabla \cdot \mu (\phi) \left(\nabla \mathbf{u} +\nabla \mathbf{u}^T \right) + G(\phi) \nabla \phi \\ 
     &&\nabla \cdot \mathbf{u}
    = 0.
    \label{eq:continuity}
\end{eqnarray}
In Eq.~\ref{eq:NS}, $\tau$ is a extra stress tensor due to the non-Newtonian rheology and $G(\phi) \nabla \phi$ represents the surface tension force.
The second term in the left-hand side represents the mass flux due to diffusion, given by $\mathbf{J} = -(\rho_1-\rho_2) M \nabla G /2$, where $\rho_1$ and $\rho_2$ are the densities of the liquid and gas phases \cite{Abels2012}.

The constitutive model for a Giesekus fluid is
\begin{eqnarray}
    &&\mathbf{\tau} + \lambda \left(\frac{\partial \tau}{\partial t} +\mathbf{u} \cdot \nabla \mathbf{\tau} -\mathbf{\tau} \nabla \mathbf{u}-  \nabla \mathbf{u}^T \mathbf{\tau} \right) + \frac{\alpha \lambda} {\mu_p} (\tau \cdot \tau)
    = \mu_ p \left(\nabla \mathbf{u} + \nabla \mathbf{u}^T \right),
    \label{eq:Giesekus}
%
  \end{eqnarray}
%
where $\lambda$, $\alpha$, $\mu_p$ are the relaxation time, the anisotropic factor, and the polymeric viscosity, respectively. 
In the single mode Giesekus model \cite{YESILATA_JNNFM200673}, the following analytical expression of shear viscosity can be derived, 
\begin{equation}
    \mu =  \mu_0 \frac{(1-f)^2}{1+(1 - 2 \alpha)f}+\mu_s.
     \label{eq:Viscosity}
\end{equation}
Here, $f, \kappa_1$ and $\kappa_2$ are functions of $\lambda$, $\alpha$ and the shear rate $\dot\gamma$. Their explicit expressions can be found in the Appendix (\ref{eq:Viscosity_f}, \ref{eq:Viscosity_k1}, \ref{eq:Viscosity_k2}).  
The total viscosity or zero-shear-rate viscosity in Eq. \ref{eq:Viscosity} is given by
\[
\mu_0 = \mu_s +\mu_p,
\]
where $\mu_s$ is the solvent viscosity of the liquid phase.
For all the performed simulations $\alpha$ and $\mu_s$ are set to 0.4955 and $0.1 \mu_0$, respectively. An example of the shear viscosity given by Eq.~\ref{eq:Viscosity} is shown in Fig.~\ref{fig:rheology} (solid black line), where the shear thinning effect is clearly observed ($\mu_0 = 1 \times 10^{-2}~\mathrm{Pa \cdot s}$ and $\mu_\infty\approx 10^{-3}$ Pa$\cdot$s). 
The phase-field mobility $M$ in Eq.~\ref{eq:PF} is set using an equivalent definition $M = 2 \eta \alpha_{d}/\sigma$, where $\alpha_{d} = 5.7 \times 10^{-6}$ is the mass diffusion constant in the literature \citep{CarlsonJFM2011}.
Cahn number $Cn = \eta/R_{0}$ represents the ratio between the diffuse interface width and the characteristic length scale and is set to $Cn = 0.008$. The equations are discretized using a finite difference method. The detailed numerical scheme and the validation of the numerical method can be found in \cite{bazesefidpar2022_pub}.

\subsection{Model of contact-line dissipation}
Contact-line motion can be modelled by imposing a variety of boundary conditions on the velocity field and the phase-field variable. See \cite{Lacis2020, lacis_pellegrino_sundin_amberg_zaleski_hess_bagheri_2022} for a thorough comparison between different boundary conditions. 
We impose a no-slip condition for $\mathbf{u}$ on the solid wall, and a non-equilibrium boundary condition~\cite{YuePF2011, CarlsonPoF2009} for $\phi$ at the contact line. The latter is given by the expression,
\begin{equation}
	-\eta \mu_{f}  \frac{\partial \phi}{\partial t} = \eta \sigma \nabla \phi \cdot \textit{\textbf{n}} - \sigma {\rm cos} (\theta _{e})g^{\prime} (\phi).
	\label{eq:muf}
\end{equation}
Here, \(\theta_{e}\) is the static contact angle and $\mu_f$ is the contact-line friction parameter, discussed earlier. The static contact angle $\theta_e$ is fixed to $52^{\circ}$ to match to the experiments for all cases. The polynomial $g(\phi) = 0.5+0.75\phi-0.25\phi^3$ rapidly shifts from $0$ (in vapor phase $\phi = -1$) to $1$ (in liquid phase $\phi = 1$). 

Note that the contact line motion is also influenced by the diffusion in \ref{eq:PF}, which depends on the interface thickness $(\eta)$ and the mobility $(M)$ parameter. Here, both $\eta$ and $M$ are constant and treated as numerical parameters
\cite{Lacis2020, lacis_pellegrino_sundin_amberg_zaleski_hess_bagheri_2022}. 
The contact-line friction is used to account for the local energy dissipation at the contact line. In \ref{eq:muf}, this dissipative force is modelled by a linear friction law, i.e. the left-hand side of \ref{eq:muf} is $\sim \mu_f U_{cl}$. The right-hand side of \ref{eq:muf} is related to 
%
Young's force ($\sim \sigma (\cos\theta_e - \cos\theta$)) that drives the wetting 
($\theta$ being the dynamic contact angle). We  note that a zero $\mu_f$ yields $\theta=\theta_e$. On the other hand,  rapid wetting of a droplet on a hydrophilic surface ($\theta\gg\theta_e$) results in a large friction force and contact-line dissipation.
The line friction parameter $\mu_f$ for water on the OSTE substrate is identified to 0.12~Pa$\cdot$s by fitting the numerical spreading curve to the experimental spreading curve~\cite{Yada_langmuir_2021}. This is explained in detail in appendix \ref{app_muf}.

\begin{table}
    \centering
    \begin{tabular}{llllccc}
    \hline
         & & $\mu_{0}$    & $\mu_{f}$ & De \\
          & & (Pa$\cdot$s) & (Pa$\cdot$s)  & ($\lambda U_{ref}/R_0)$   
    \\ \hline
          W & Newtonian & $1.0 \times 10^{-3}$  & 0.12 & -    \\
  Gi:1\hspace{5mm}  & Giesekus\hspace{5mm}  & $1.0 \times 10^{-3}$ \hspace{5mm}  & 0.12 \hspace{5mm} & 20   \\
     Gi:2  & Giesekus & $1.0 \times 10^{-3}$  & 0.06 & 20 \\
      Gi:3  & Giesekus & $1.0 \times 10^{-3}$  & 0.24& 20  \\
      Gi:4 & Giesekus & $1.0 \times 10^{-3}$  & 0.36& 20    \\
     Gi:5  & Giesekus & $2.0 \times 10^{-3}$  & 0.12& 20  \\
      Gi:6  & Giesekus & $5.0 \times 10^{-3}$  & 0.12& 20   \\
      Gi:7  & Giesekus & $1.0 \times 10^{-2}$  & 0.12& 20   \\
     Gi:8  & Giesekus & $1.0 \times 10^{-3}$  & 0.12& 5  \\
      Gi:9  & Giesekus & $1.0 \times 10^{-3}$  & 0.12& 50 \\
     N:1 & Newtonian & $1.0 \times 10^{-3}$  & 0.06 & -    \\ 
     N:2 & Newtonian & $1.0 \times 10^{-3}$  & 0.24 & - \\ 
     N:3 & Newtonian & $1.0 \times 10^{-3}$  & 0.36 & - \\ 
     N:4 & Newtonian & $2.0 \times 10^{-3}$  & 0.12 & -   \\ 
     N:5 & Newtonian & $5.0 \times 10^{-3}$  & 0.12 & -\\ 
     N:6 & Newtonian & $1.0 \times 10^{-2}$  & 0.12 & -  \\ 
    \hline
    \end{tabular}
    \caption{List of simulation input parameters. $Cn = \eta/R_0$ =0.008 and Pe =$U_{ref}R_0 /\alpha_d$ = 33.3  are fixed for all cases. The anisotropic factor $\alpha$ and $\mu_s$ are set to 0.4955 and $0.1 \mu_0$ for all cases.  
    }
   \label{table:simulations}
\end{table}

\begin{figure}
    \centering
      \includegraphics[width=0.6\textwidth]{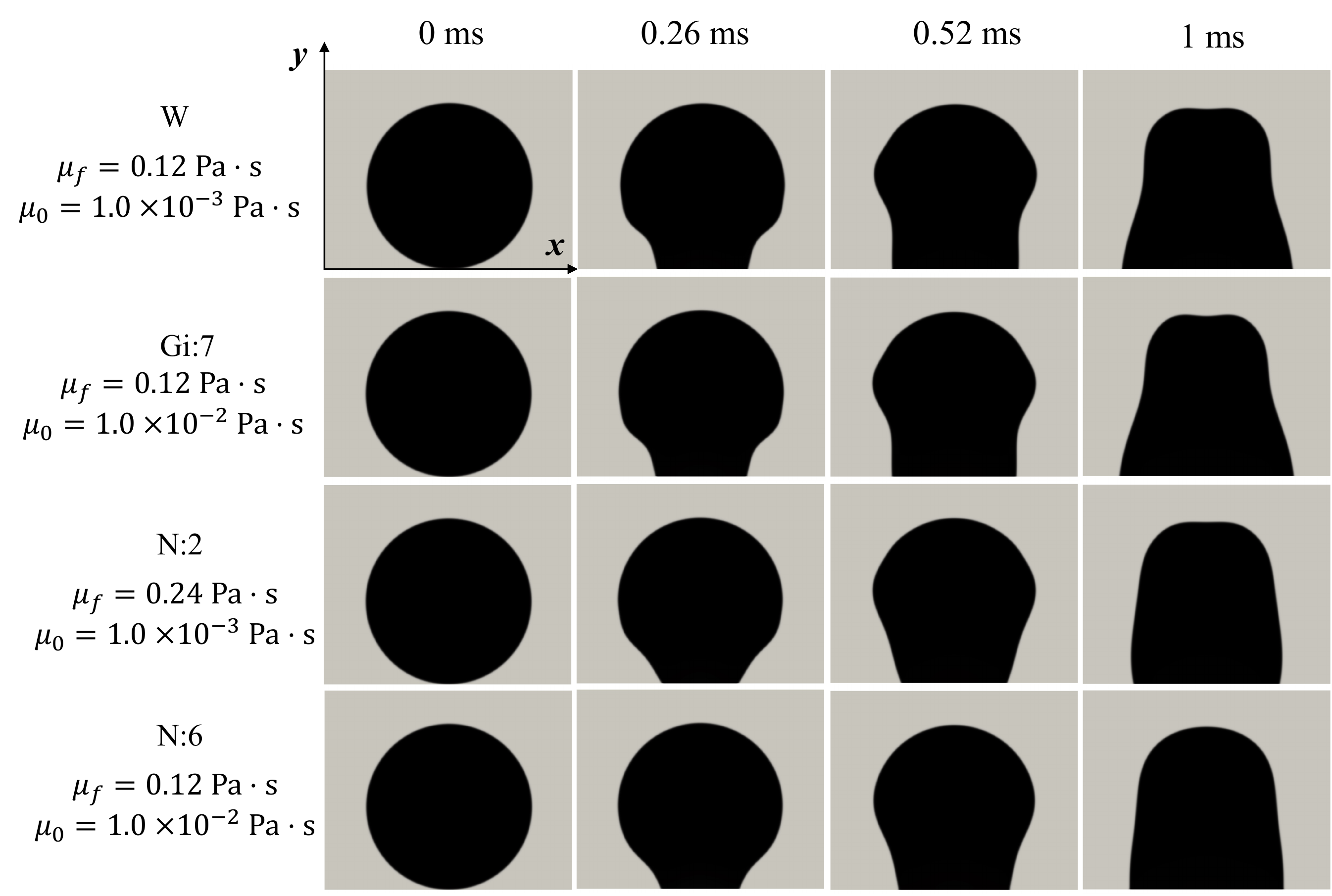}
    \caption{
   Snapshots from simulations. The black and grey phases indicate liquid and air phases, respectively. Spatial coordinates are indicated in the top-left frame.
     }
    \label{fig:Numsnaps}
\end{figure}

\begin{figure}
    \centering
     \includegraphics[width=0.3\textwidth]{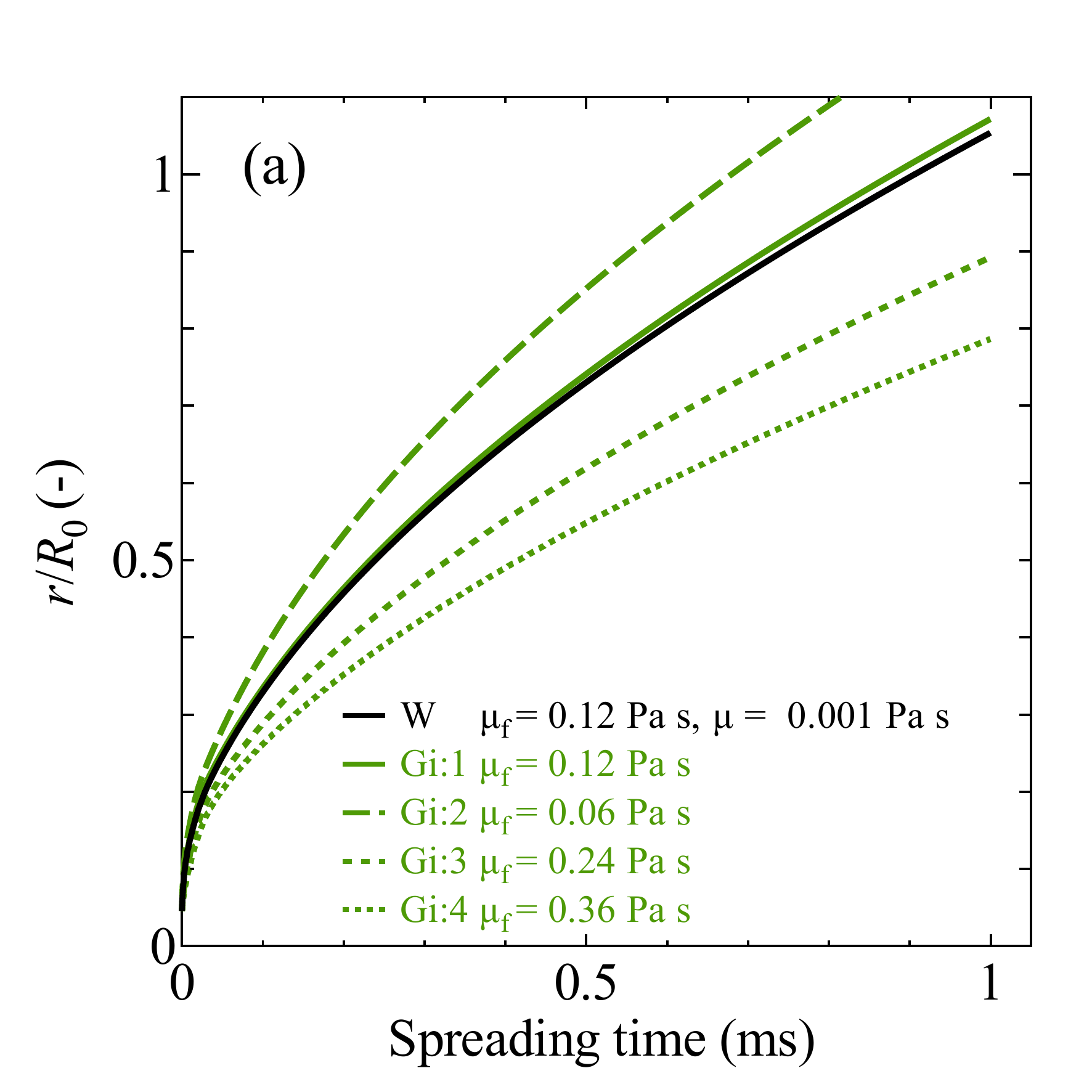}
       \includegraphics[width=0.3\textwidth]{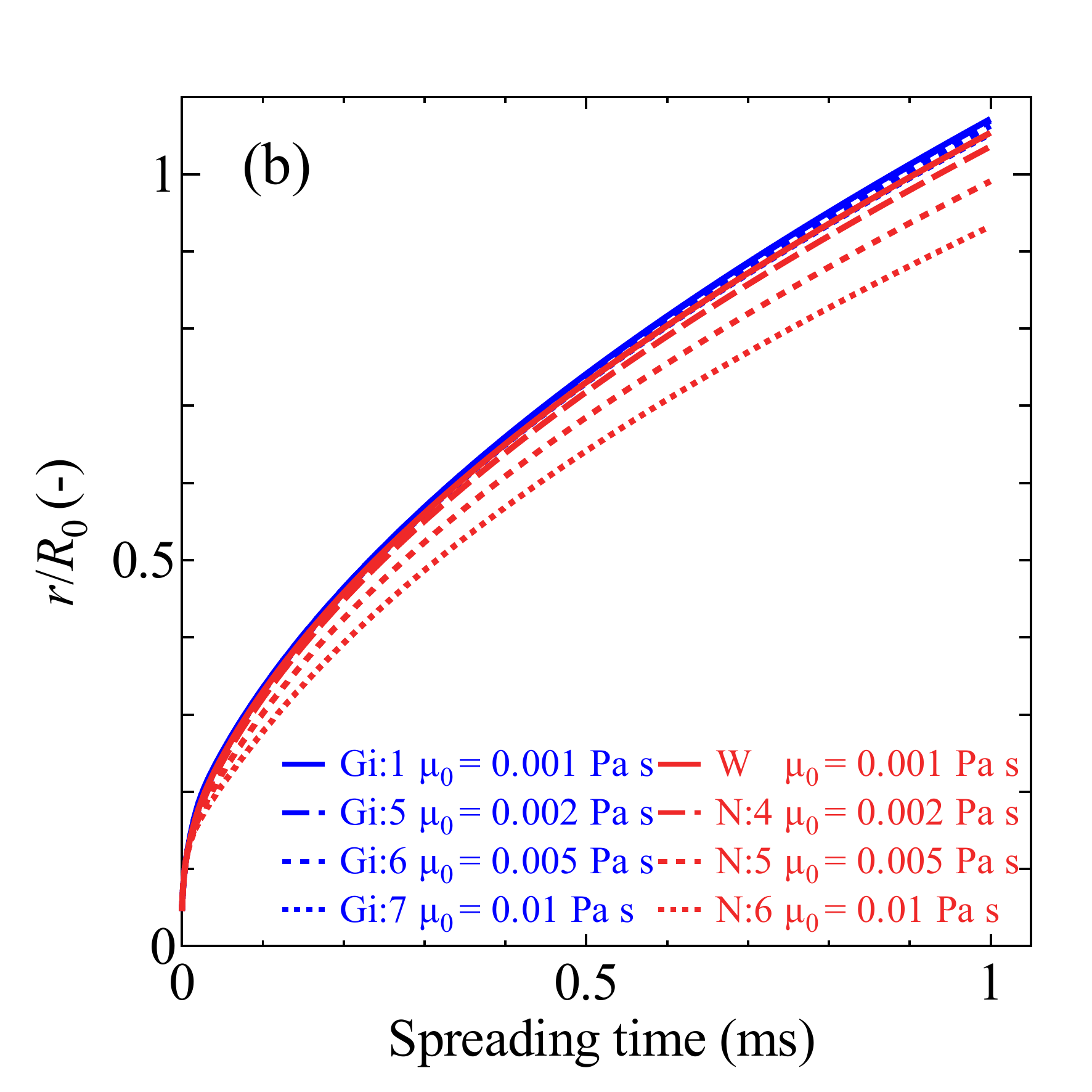}
       \includegraphics[width=0.35\textwidth]{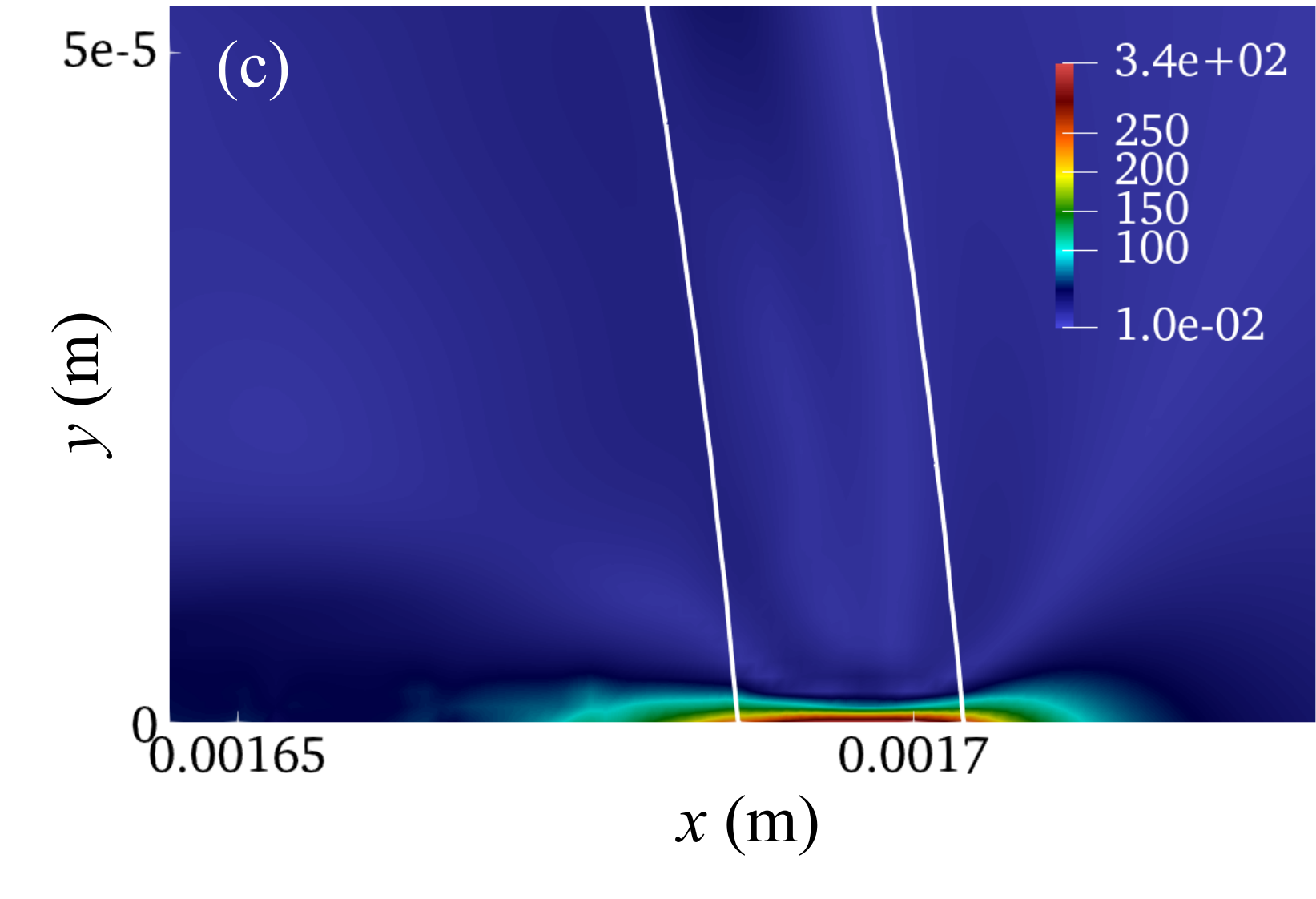}
    \caption{
   (a) Spreading radius with different line friction parameter with $\mu_0 = 1 \times 10^{-3}~\mathrm{Pa \cdot s}$. (b) Comparison between Newtonian (red curves) and Giesekus droplet (blue curves) with different zero-shear viscosity with $\mu_f$ = 0.12~Pa$\cdot$s. 
   (c) Normalized shear rate distribution $T_f \dot \gamma = T_f \sqrt{(\gamma_{xx}^2+2\gamma_{xy}^2+\gamma_{yy}^2)/2}$ for Gi:1 at $t$ = 1 ms. The white lines in (c) are the contour lines of $\phi = \pm 0.9$. The interface is moving from left to right.
     }
    \label{fig:Numcurve}
\end{figure}

\subsection{Resistance from viscoelastic stress and contact-line friction}

We conduct a parametric study by varying the total viscosity $\mu_0$, the line friction parameter $\mu_f$, and the relaxation time of the polymers $\lambda$.
The simulation cases and input parameters are summarized in table \ref{table:simulations}.
The relaxation time is listed using the Deborah number
\[
De = \frac{\lambda U_{ref}}{R_0}.
\]
Here, $U_{ref} = \sqrt{\sigma/\rho R_0}$ is the capillary-inertial velocity scale. The configuration denoted by "W" models a water droplet, where both the viscosity and friction factor has been matched with experiments (see Appendix \ref{app_muf}). The configurations denoted by Gi:1 to Gi:9 are Giesekus droplets with varying $\mu_0, \mu_f$ and $\lambda$. Finally, the configurations from N:1 to N:6 model Newtonian droplets with varying $\mu_0$ and $\mu_f$. 
All the simulations are two-dimensional and initialized with a droplet in contact with the solid surface as shown in the first column of Fig.~\ref{fig:Numsnaps}. The radius of the droplet $R_0$ is set to 0.5~mm to match to the experiments. Surface tension $\sigma$ and the liquid density $\rho$ are 0.072 N/m, 1 $\times 10^3 \mathrm{kg/m^3}$, respectively. The air viscosity and density are set to 1.6 $\times 10^{-5}$~Pa$\cdot$s and 1.2~$\mathrm{kg/m^3}$, respectively. 

Figure ~\ref{fig:Numsnaps} shows snapshots of a water droplet (W, first row), a Giesekus droplet (Gi:7, second row), and two Newtonian droplets (N:2 and N:6 in 3rd and 4th rows). This figure can be compared to the experimental snapshots in Fig.~\ref{fig:snaps}. We again observe that the droplet shapes of water and non-Newtonian liquid are very similar in the first millisecond. This is despite the fact that zero-shear rate viscosity, $\mu_0$, is an order of magnitude larger for case Gi:7 compared to the viscosity of water. Here, we have assumed that the energy dissipation at the contact line is not modified by the polymers, and thus chosen $\mu_f=0.12$ Pa$\cdot$ s for both droplets. In contrast, when we increase the bulk viscosity (N:6) or the contact-line friction (N:2) of a Newtonian droplet, we notice a different droplet shape after one millisecond. By comparing these snapshots to their experimental counterparts in figure \ref{fig:snaps}, we can confirm that our numerical treatment captures the physics of Newtonian and non-Newtonian droplets during spreading.

Our numerical model allows for independently varying the amount of viscous and contact-line dissipation during rapid spreading for Newtonian and non-Newtonian droplets. Figure~\ref{fig:Numcurve}(a) shows the spreading radius of Giesekus droplets (Gi:1 to Gi:4) with different line friction parameter and fixed total viscosity ($\mu_0 = 1 \times 10^{-3}~\mathrm{Pa \cdot s}$). The spreading rate of the water is shown with a black solid line. We observe that the line friction critically determines the spreading radius. As the line friction parameter is increased, the spreading rate slows down.

Figure~\ref{fig:Numcurve}(b) compares spreading of the Giesekus and Newtonian droplets with different total viscosity $\mu_0$, but fixed contact-line friction ($\mu_f$ = 0.12~Pa$\cdot$s). Giesekus droplets (Gi:1, Gi:5 to Gi:7) are shown in blue color and the Newtonian (W, N:4 to N:6) in red color.
The spreading of the Giesekus droplet is insensitive to the total viscosity even when it is ten times higher ($\mu_0 = 1 \times 10^{-2}~\mathrm{Pa \cdot s}$). 
In contrast, viscosity influences spreading rate of Newtonian droplets (red curves). 
The difference between the Newtonian and Giesekus droplets lies in the shear-thinning effect.
As seen in Fig.~\ref{fig:Numcurve}(c), the shear rate near the contact line is high compared to the interior part of the liquid.
The effective viscosity of the Giesekus droplets seems to be the viscosity at a high-shear rate, which is much smaller than the total viscosity.  In our numerical framework, the high-shear viscosity is set to $0.1 \mu_0$.
%
%

\begin{figure}
    \centering
           \includegraphics[width=0.35\textwidth]{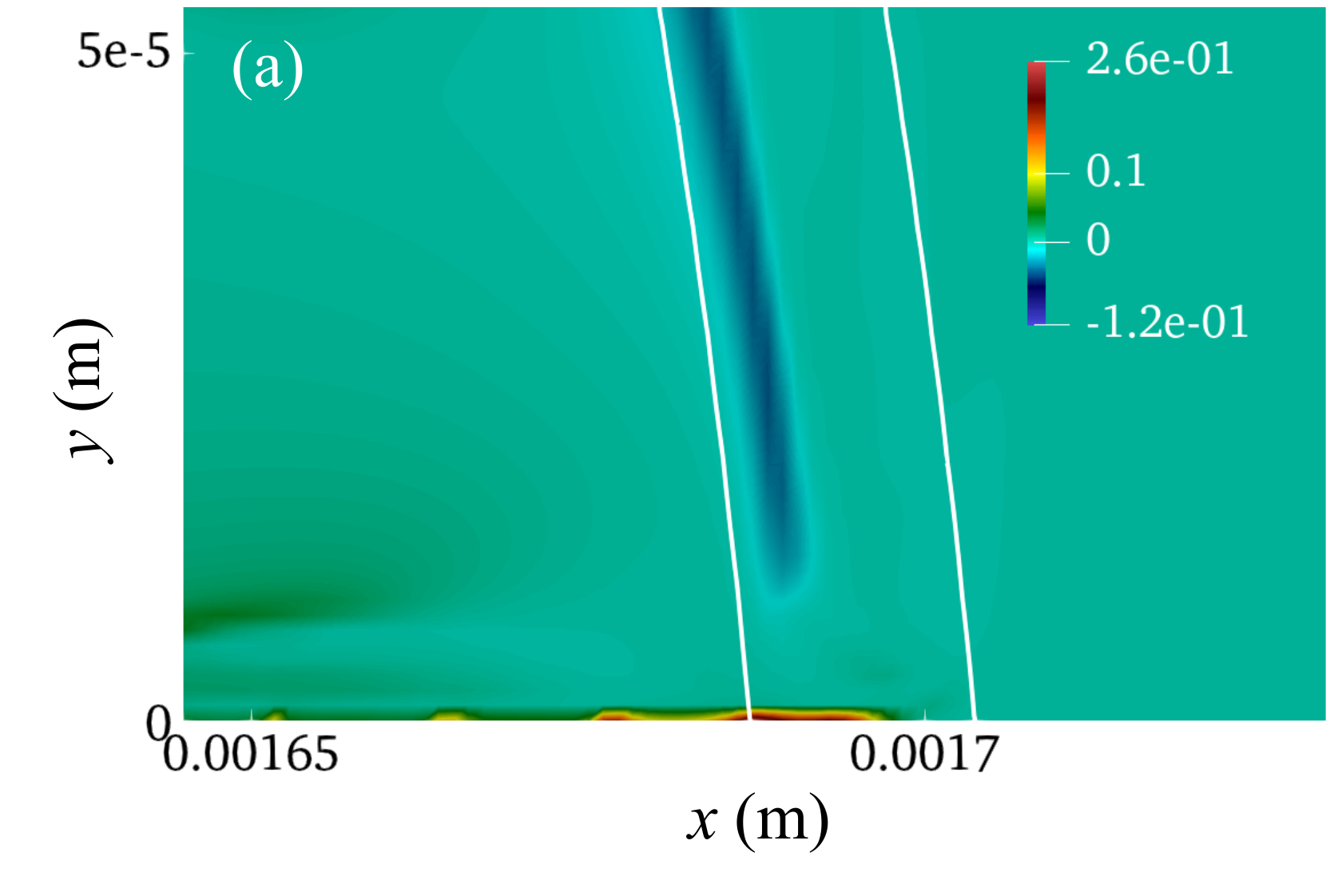}
    \includegraphics[width=0.3\textwidth]{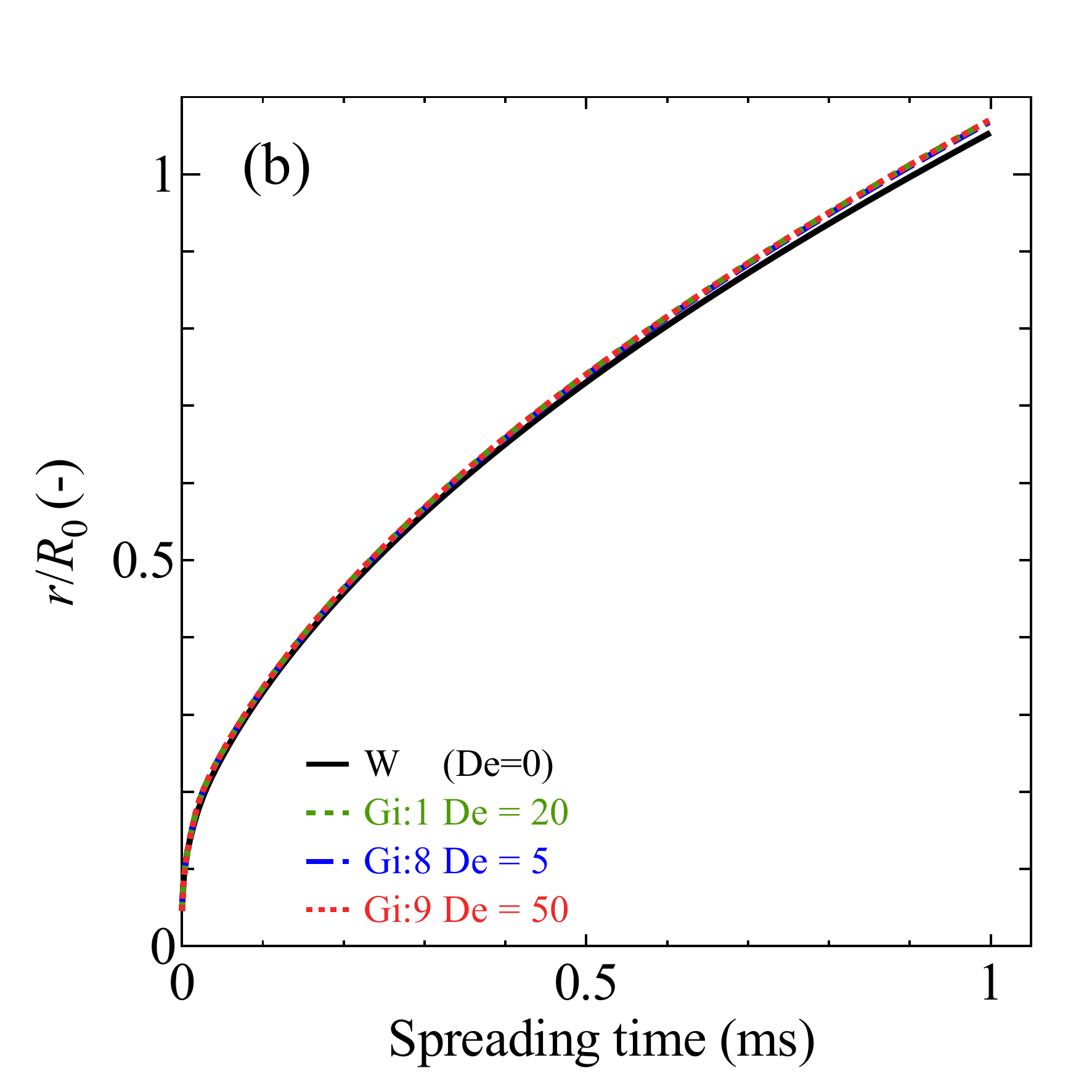}
    \includegraphics[width=0.3\textwidth]{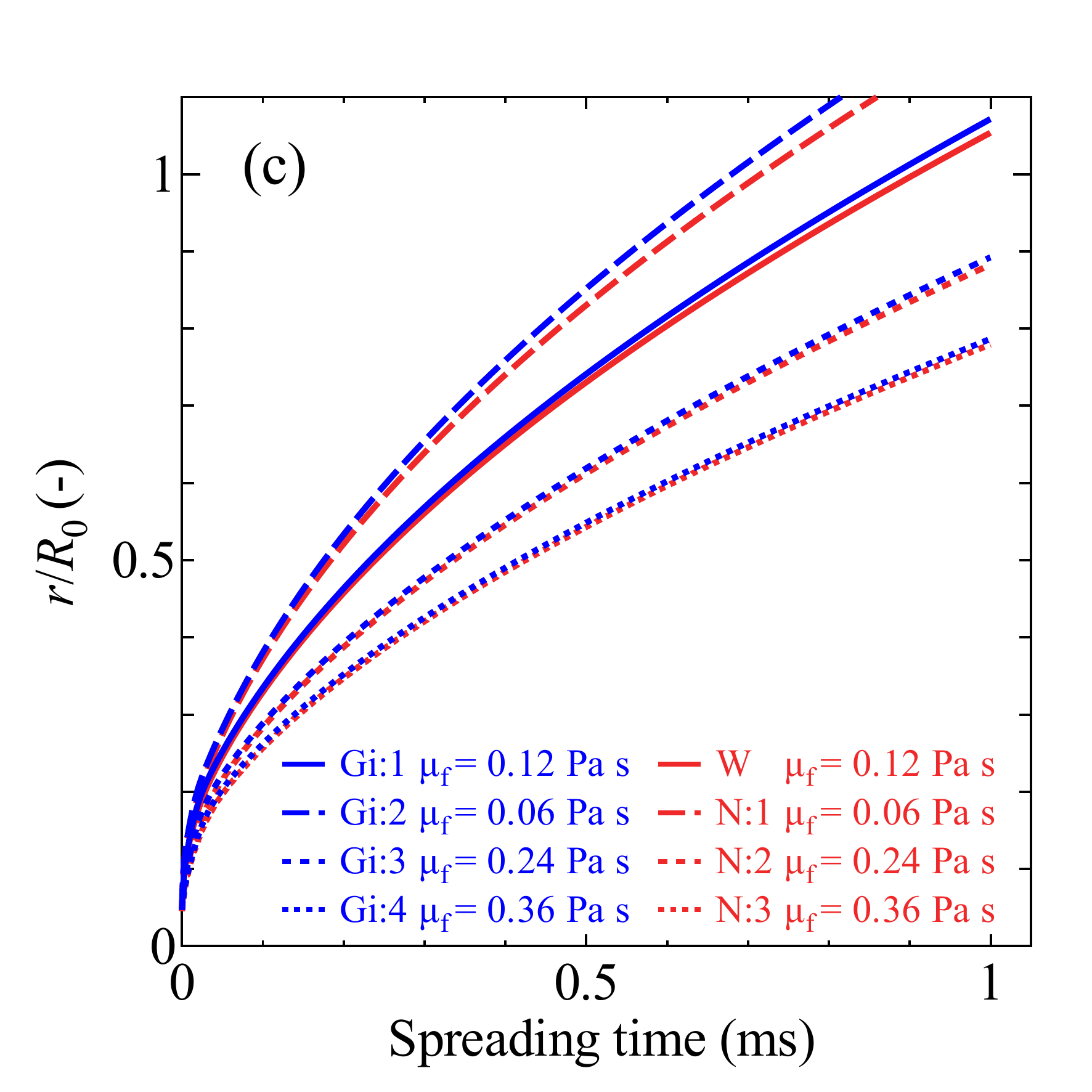}
    \caption{
    (a) The first normal stress difference $(\tau_{yy}-\tau_{xx})/\rho U_{ref}^2$ of Gi:1 at $t$ = 1 ms. The white lines in (d, e) are the contour lines of $\phi = \pm 0.9$. The interface is moving from left to right.
   (b) Spreading radius with different Deborah number with $\mu_0 = 1 \times 10^{-3}~\mathrm{Pa \cdot s}$ and $\mu_f = 0.12$~Pa$\cdot$s. 
   (c) Comparison between Newtonian (red) and Giesekus droplet (blue) with different line friction parameters $\mu_0 = 1 \times 10^{-3}~\mathrm{Pa \cdot s}$. 
     }
    \label{fig:Stress}
\end{figure}
The polymeric fluids in the numerical study are not only viscous but also elastic. The first normal stress difference $(\tau_{yy}-\tau_{xx})/\rho U_{ref}^2$ is shown in Fig.~\ref{fig:Stress}(a). The polymeric stress is concentrated in the vicinity of the contact line. This agrees with earlier work \citep{WangPRE2015, YUE_JNNFM_2012}. 
The magnitude of the stress is an order of magnitude smaller than the characteristic fluid pressure ($\sim\rho U_{ref}^2 = \sigma/R_0 $) even near the contact line and is negligible in the rest of the droplet.
This implies that 
the fluid elasticity does not contribute significantly to the spreading speed. 
Moreover, $De$ indicates the degree of elasticity.
The spreading is insensitive to the Deborah number in the range of 5 - 50 (see Fig.~\ref{fig:Stress}b). This also implies that the effect of the elasticity is small. 

\citet{WangPRE2015} reported that a Giesekus droplet may spread faster than a Newtonian droplet with the same total viscosity of the droplet $\mu_{0}$. This is not observed in our numerical simulations. The apparent discrepancy is due to the different boundary conditions imposed at the contact line for the Cahn-Hilliard equation. \citet{WangPRE2015} employed the equilibrium condition, i.e, $\mu_f =0$ while in this study the non-equilibrium boundary condition $\mu_f > 0$ is imposed. As seen in Fig.~\ref{fig:Stress}(c), for a smaller $\mu_f$, we also observed a slightly faster spreading of the Giesekus droplet than a Newtonian droplet with the same $\mu_0$. Particularly, for $\mu_f =0.06 $~Pa$\cdot$s, the Giesekus droplet spreads faster than the Newtonian droplet.
The difference between the Giesekus droplet and Newtonian droplet with the same total viscosity becomes smaller as the contact line friction increases.

\section{Droplets spreading quantified in terms of $Oh$-$Oh_f$ map}
The numerical results of the previous sections clarified how the spreading curves change with respect to contact line friction (Fig.~\ref{fig:Numcurve}a), the total viscosity (Fig.~\ref{fig:Numcurve}b), and the relaxation times (Fig.~\ref{fig:Stress}b). We found that for shear-thinning fluids only the contact-line friction seems to modify the spreading curves in the rapid regime. In other words, the relevant sources of wetting resistance are either inertial forces, the contact line friction, or both.
Moreover, the similar spreading curves of water and shear-thinning fluids observed in experiments (Fig.~\ref{fig:Data}) indicate that the contact-line friction is not modified by addition of polymers. This means that we can assume $\mu_f\approx 0.12$ Pa$\cdot$s for all the non-Newtonian droplets.
These insights can now be gathered in a parameter map spanned by different Ohnesorge numbers.

\subsection{Ohnesorge numbers}
The Ohnesorge number $Oh$ based on $\mu_\infty$ is defined as 
\[
Oh = \frac{\mu_{\infty}}{\sqrt{\rho \sigma R_0}},
\]
where $\rho$ is the density of the solution and $R_0$ is the initial radius of the droplet.  The Ohnesorge numbers, $Oh\ll 1$, reported in Tab.~\ref{table:merged} indicate a small influence of high-shear-rate viscosity compared to inertia near the contact line for all the investigated liquids. As shown in the previous section, $\mu_{\infty}$ is the relevant viscosity near the contact line where the shear rate is high. The conventional Ohnesorge number remains significantly smaller than unity even if $\mu_0$ is increased because $\mu_\infty$ remains small.

The convectional $Oh$ alone does not determine the spreading rate, since line-friction may provide additional resistance to wetting.
For example, in Tab.~\ref{table:merged}, we note that 60$\%$ glycerol solution has $Oh_{\infty}=6.2\times 10^{-2}$, which is very close to corresponding values for FP 500 ppm, PEO 500 ppm, and XG 2000 ppm. However, figure~\ref{fig:Data} clearly demonstrates the slower spreading rate of the 60$\%$ glycerol solution compared to the non-Newtonian liquids.
The relative importance of the line friction parameter to inertia is given by the Ohnesorge number based on $\mu_f$
\[
Oh_{f} = \frac{\mu_{f}}{\sqrt{\rho \sigma R_0}}.
\]
As shown in Table \ref{table:merged}, the line friction Ohnesorge number is order one for a water droplet and significantly larger than the conventional Ohnesorge number. This implies that the contact line friction contributes to the energy dissipation of the droplet much more than the viscosity. A significant contact-line dissipation also takes place in our polymeric solutions. 

Based on our observations from Fig.~\ref{fig:Data} however, $Oh_{f}$ is independent of the polymer concentration. In other words, the polymer chains seem to not influence the energy dissipation at the contact line for shear-thinning fluids.
Table \ref{table:merged} also shows that the line friction parameter increases with increasing glycerol concentration for $60\%$ and $80\%$ concentration and it exceeds $Oh_f > 1$.

\begin{figure}
    \centering
    \includegraphics[width=0.5\textwidth]{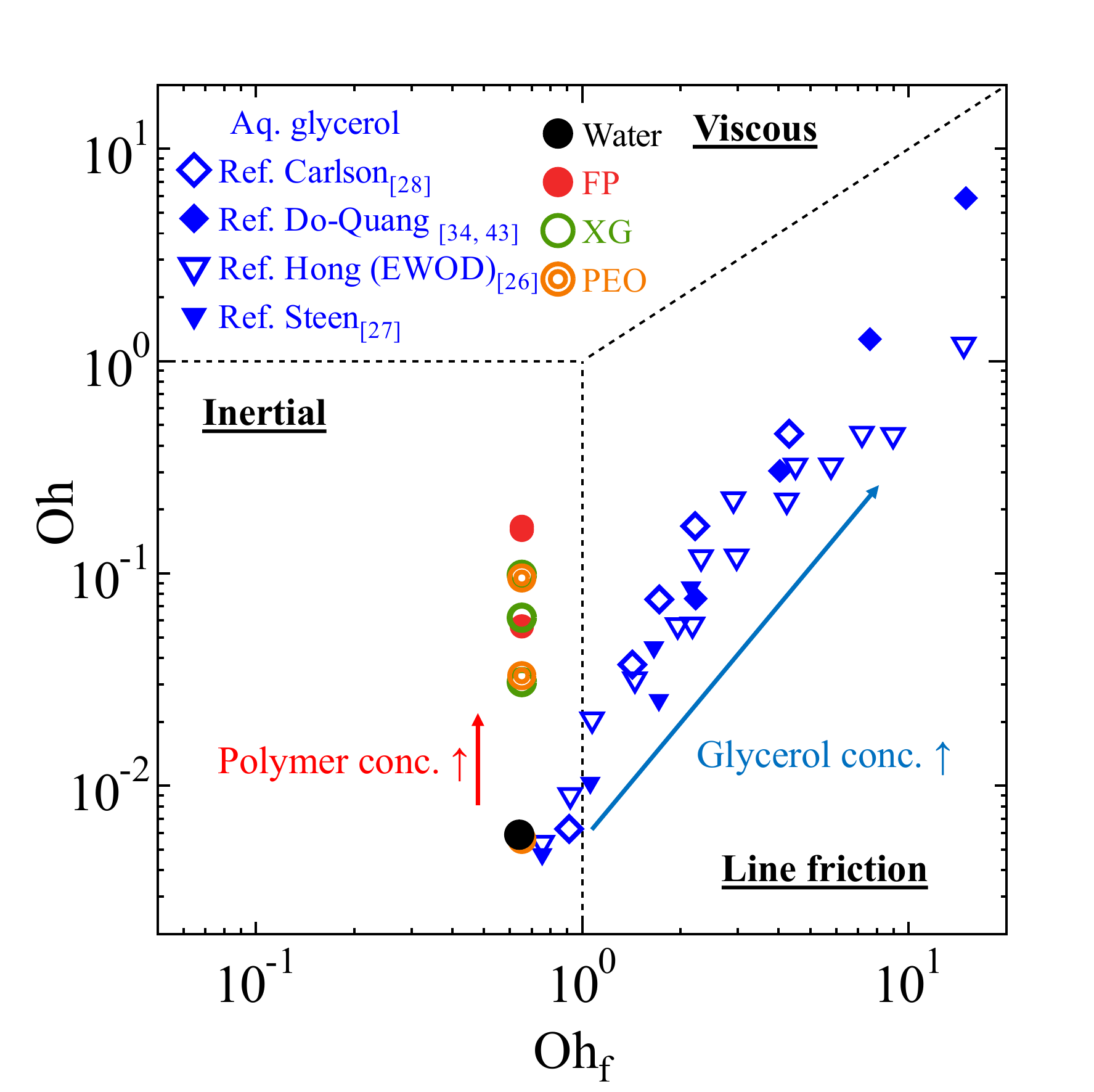}
    \caption{
    $Oh_{f}$ - $Oh$ map for a droplet with the initial radius of 0.5 mm. The viscosity-line friction parameter data on silanized surfaces \cite{CarlsonPRE2012} and glass surfaces \cite{Do-Quang2015, Eddi2013}.
    The viscosity-line friction data of aqueous glycerol from oscillatory droplet measurements \cite{Steen2020} and electrowetting on dielectric layers(EWOD) \cite{Hong2013Langmuir} are also plotted. The shear viscosity at the largest shear rate are employed for FLOPAM (FP), Xanthan gum (XG), and PEO solutions (TABLE \ref{table:merged}). 
     }
    \label{fig:OhOhf}
\end{figure}
It is instructive to represent our results together with other existing data of Newtonian droplets in a parameter space spanned by $Oh$ and $Oh_f$. This map has previously been used to cover different spreading regimes in rapid wetting \cite{Do-Quang2015}.
Figure~\ref{fig:OhOhf} shows the parameter map and the regimes where the spreading rate is primarily resisted by inertial forces, viscous forces, or the contact-line friction.
For the water droplet considered here ($R_0$ = 0.5 mm, $\mu = 1\times 10^{-3}~\mathrm{Pa\cdot s}$, and $\mu_f = 0.12~\mathrm{Pa\cdot s}$), we have $Oh_f = 0.63$ and $Oh = 5.3 \times 10^{-3}$.
The water droplet is thus in the inertial regime, but since $Oh_f \approx 1$, one may expect a contribution also from the contact line friction.
Glycerol droplets are shown with blue symbols.  As their concentration increases, both the line friction parameter and the viscosity increase \cite{Do-Quang2015, WangSCIREP2015, Vo2018,Steen2020, Hong2013Langmuir}. %
Therefore, the glycerol solutions are in the line friction governed region $Oh_f >1$. 
Finally, polymers in pure water are shown in red, green, and yellow colors. 
The additive polymers do not increase the line friction and the high-shear-rate viscosity remains sufficiently small so that the spreading of the polymeric solutions is in the inertial regime.

The 5000 ppm Xanthan gum/FLOPAM solutions show slightly slower spreading in the experiments (see red/green triangles in Fig.~\ref{fig:Data}a, b) since the solutions may reach the regime $Oh \approx 0.1$ where viscosity starts to play a role in the spreading.  

The diagram provides insight into how the rapid spreading of shear-thinning liquids can be controlled. When the Ohnesorge number based on the solvent viscosity is small ($Oh \ll 1$), additional polymeric viscosity is not a critical factor to control a rapid spreading. Rather, modifying the solid surface property is more effective when $Oh \ll 1$ and $Oh_f \approx 1$. A possible way to enhance the contact line friction is to introduce roughness. On a non-smooth surface, an effective line friction parameter $\mu_{f eff}  = S \mu_f$ can be defined \cite{WangSCIREP2015, Lee2019, Yada2019} where $S$ is a measure of surface roughness. This roughness would critically retard the spreading and the spreading speed scales as $\sim \sigma/S \mu_f$ \cite{WangSCIREP2015, Lee2019, Yada2019}.

\begin{figure}
    \centering
    \includegraphics[width=0.5\textwidth]{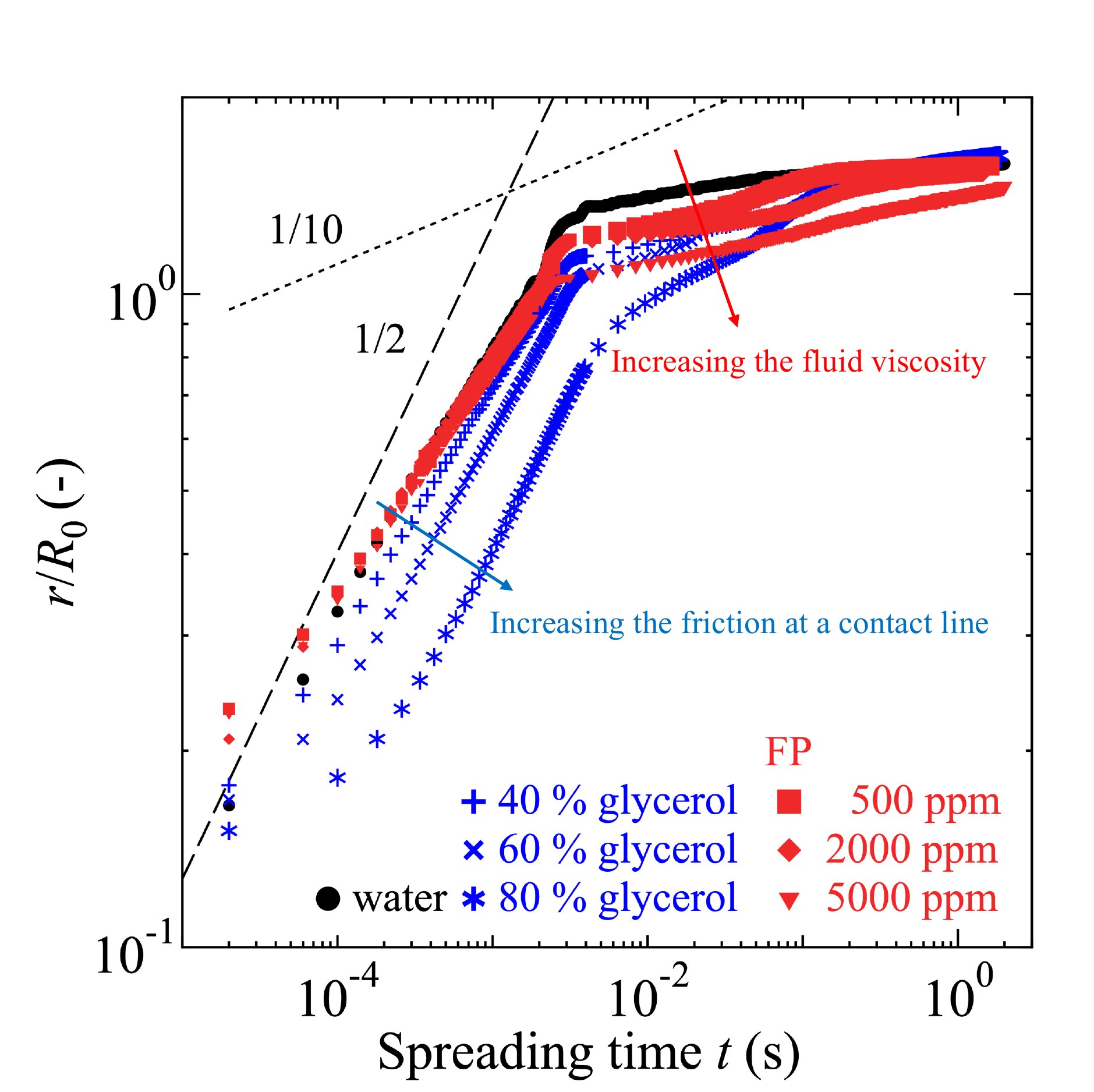}
    \caption{Spreading curves over four time decades $t$. }
    \label{fig:Late}
\end{figure}

\section{Discussion}

Figure~\ref{fig:Late} shows the spreading radius over four time decades for aqueous glycerol solutions and shear-thinning solutions. Note that all fluids have similar static contact angles and therefore, the terminal spreading radius is nearly the same. 
Dashed lines in the figure show $r/R_0 \propto t^{1/2}$ and $r/R_0 \propto t^{1/10}$. 
The latter spreading rate follows Tanner's law \cite{Tanner1979}, where viscous dissipation is the dominating source of resistance to capillary-driven wetting.
In the late spreading, a difference with respect to polymer concentrations is observed; solutions with higher concentration spread slower after $t \approx 10^{-3}$~s. This agrees with the slower late spreading of shear-thinning fluids observed by Rafai \textit{et al.}~\cite{Rafai2004}.

We can again confirm from Fig.~\ref{fig:Late} that the shear-thinning solutions exhibit very similar spreading curves to water regardless of the polymer concentration until $t \approx 10^{-3}$ s. In this inertial regime, the spreading rate follows a time scale set by a balance between the rate of change of kinetic energy and the driving surface tension, i.e.
$
t_i \sim \left ({\rho R_0^3}/{\sigma}\right)^{1/2}.
$
%
This is in agreement with \citet{Ambre_Snoeijer_Arxiv}, who very recently reported that the contact line speed is only slightly decelerated by the PEO polymers in water up to 2.0wt $\%$. The authors concluded that the rapid spreading behavior is "inertial" in nature because of the spreading exponent $\approx$ 1/2.

Figure ~\ref{fig:Late} also re-confirm that the aqueous glycerol solutions spread slower than the water in the initial regime and exhibit a longer rapid wetting regime (up to $t \approx 10^{-2}$~s). As the glycerol solution increases, the initial spreading is slowed down (marked by the blue arrow in Fig.~\ref{fig:Late}). 
The increasingly more concentrated glycerol solutions are in the ``line-friction regime'', where the time scale is set by $t_f=\mu_f R_0/\sigma$ \cite{Do-Quang2015, CarlsonPRE2012}. 
We also note that the spreading radius of the glycerol solutions all follow $r/R_0 \propto t^{-1/2}$, with a multiplicative prefactor that decreases
with increasing concentration, i.e. the line friction parameter. This confirms the scaling of spreading in the line friction regime, $r/R_0 \sim ( \sigma t / \mu_f R_0 )^{1/2}$, as proposed in [28].
%

Our interpretation is that two features are responsible for polymer solutions remaining in the inertial regime regardless of the polymer concentration; i) polymers do not modify the contact line friction; ii) the relevant high-shear-rate viscosity near the contact-line is small. The first point means that $Oh_f$ is independent of polymer concentration, while the latter point means that $Oh<1$. Indeed, both these points need to be satisfied to explain the experimental result  . If the second point was a sufficient condition, then the addition of polymers would result in a side-ways (horizontal) movement in the $Oh_f-Oh$ map, and thus -- similar to glycerol solutions -- a slower spreading rate in the rapid wetting regime.

Further studies are required to establish the reason for the independence of line-friction from polymers. 
One plausible reason is that  
the polymers migrate from the contact line. As seen in Fig.~\ref{fig:Numcurve}(c), the shear rate is very high near the contact line.
Revisiting the molecular migration theory \cite{Ma2005, HAN2013}, the polymer molecules subjected to strong shear move away from the contact line region to the bulk, and the polymer concentration at the contact line is notably lower. 
Han \textit{et al.} \cite{HAN2013} estimated that the depletion length for heavy polyisobutylene molecules ($M_w \approx 4.2 \times 10^6$) is a few micrometers but lighter polyisobutylene molecules ($M_w \approx 5 \times 10^4$) exhibit thinner depletion layer ($\ll 1~\mathrm{\mu m}$), based on the theory by \citet{Ma2005}.
Fang \textit{et al.}~\cite{FangJR2005} observed a DNA-depleted layer up to $\approx 2-3~\mu m$ on a glass substrate, which is about one-third of the contour length of the DNA molecules. 
These studies suggest that polymer concentration is low near the contact line.
%

A second reason can be the scale separation in both space and time between water molecules ($M_w$ = 18) and the polymer molecules ($M_w > 10^6$). 
The polymers may simply be too large to fit in the wedge that is formed between the interface and the solid. Indeed, it is the fluid in this wedge that partially determines the local contact-line friction. Moreover, the time scales related to adsorption and transport of polymers may be larger than the rapid initial movement of the interface.
%
Since the polymers in this study are hydrophilic, these may be slightly depleted at the water-air interface. In an equilibrium condition, the polymers are more likely to be surrounded by water than in the water-air interface. 
Therefore, adsorption or molecular motion to the solid-water interface after the droplet makes contact with the solid could be critical for the population of the polymers near the contact line.
However, adsorption is likely to be limited during the rapid spreading time scale. The relevant non-dimensional number is the Damköhler number $ Da = R_0 /U_{ref} \tau_{ad}$ which relates the advection timescale to the adsorption timescale  $\tau_{ad}$. For example, a typical adsorption time for PEO molecules to a solid surface is a few minutes ($\approx 10^2$~s) \cite{Fu_MacroMolecules_1998, Dijt_Macromolecules_1994, Stuart_Annual_review_1996} and this leads to $Da \ll 1$. Therefore, adsorption is expected to be negligible during the rapid spreading. 

\section{Concluding remarks}

In this work, we have investigated the rapid spreading of droplets of aqueous glycerol and  polymer solutions (dilute aqueous polyacrylamide, Xanthan
gum, and polyethylene oxide (PEO)). We demonstrated that polymer solutions initially spread similarly to solvent water, regardless of the polymer concentration. In contrast, aqueous  solutions show significantly slower spreading rate than water as the glycerol concentration is increased.  
This is the case even though the magnitude of bulk internal stresses is comparable between the polymer solutions and very viscous aqueous glycerol solutions. 
%
We have used numerical simulations to decouple the dissipation stemming from viscous forces and from contact-line friction forces. We showed that the behavior of experimental spreading curves can be re-produced numerically if the contact-line friction of the non-Newtonian solutions is equal to that of pure solvent (water). In other words, the dissipation at the contact line is not modified by the presence of polymers, even at relatively high concentrations. 
We have extended the regular $Oh_f$-$Oh$ map of Newtonian fluids with polymeric solutions. A change in polymer concentration results in a vertical movement in the map. Substrate properties are thus the only way to increase contact-line friction (and thus $Oh_f$), and thus a horizontal movement in the $Oh_f$-$Oh$ map. This provides new insight into how one may control the rapid wetting of non-Newtonian fluids.
Finally, we have discussed possible reasons for the polymer-independent contact-line dissipation, which remain to be more precisely characterized in future studies.

The authors would like to thank Prof. Wouter van der Wijngaart at KTH for providing us a surface preparation environment. We also thank Prof. Per Claesson at KTH for useful discussions.

\appendix
\section{Cahn-Hilliard-Navier-Stokes equations with Giesekus constitutive model}\label{app}

In the single mode Giesekus model \cite{YESILATA_JNNFM200673}, the analytical expression of shear viscosity reads
\begin{equation}
    \mu =  \mu_0 \frac{(1-f)^2}{1+(1 - 2 \alpha)f}+\mu_s ,
     \label{eq:app:Viscosity}
\end{equation}
where
\begin{equation}
  f = \frac{1-\sqrt{(\kappa_1 -1 )/\kappa_2}}{1+(1-2 \alpha \sqrt{(\kappa_1 -1 )/\kappa_2})},
  \label{eq:Viscosity_f}
\end{equation}
\begin{equation}
  \kappa_1 = \sqrt{1+16\alpha (1-\alpha)(\lambda  \dot \gamma )^2 },
  \label{eq:Viscosity_k1}
\end{equation}
\begin{equation}
  \kappa_2 = 8 \alpha (1-\alpha)(\lambda  \dot \gamma )^2.
  \label{eq:Viscosity_k2}
\end{equation}

\section{Fitting of the contact line friction parameter }\label{app_muf}
The contact line friction parameter of water on a OSTE surface is estimated by fitting the numerical spreading curves to the experiments. Equations.~\ref{eq:PF},\ref{eq:NS}, \ref{eq:continuity}, \ref{eq:muf} are solved with finite element method, using in-house software ``FemLego'' in an axi-symmetric geometry.
FemLego is an adaptive finite element toolbox where weak formulations of partial differential equations are defined on a MAPLE worksheet\cite{AMBERG1999257}. The detailed methods are also found in Ref.~\cite{CarlsonJFM2011, Yada_langmuir_2021} 
We have employed the axi-symmetric simulations to estimate the line friction parameter since the axi-symmetric geometry is closer to the experiments. The estimation of the line friction parameters for aqueous glycerol solutions (TABLE \ref{table:merged}) follows this procedure as well.

Figure \ref{fig:A1} shows the spreading curves of numerical simulations and experiments. The numerical curve with $\mu_f =0$ overestimates the spreading speed (grey curve in Fig.~\ref{fig:A1}). 
The numerical curve with $\mu_f =0.12 $~Pa$\cdot$s (black curve) is matched well with the experiments until 0.5~ms. The deviation from experiments after 0.5~ms can be attributed to the presence of the needle in the experiments.
We also compared the axi-symmetric simulations and two-dimensional geometry in this paper (W in TABLE\ref{table:simulations}). The spreading curve also collapses on the experiments until 0.4~ms and starts to deviate due to the difference in the geometry. 

\renewcommand{\thefigure}{A\arabic{figure}} 
\setcounter{figure}{0}

\begin{figure}
    \centering
    \includegraphics[width=0.35\textwidth]{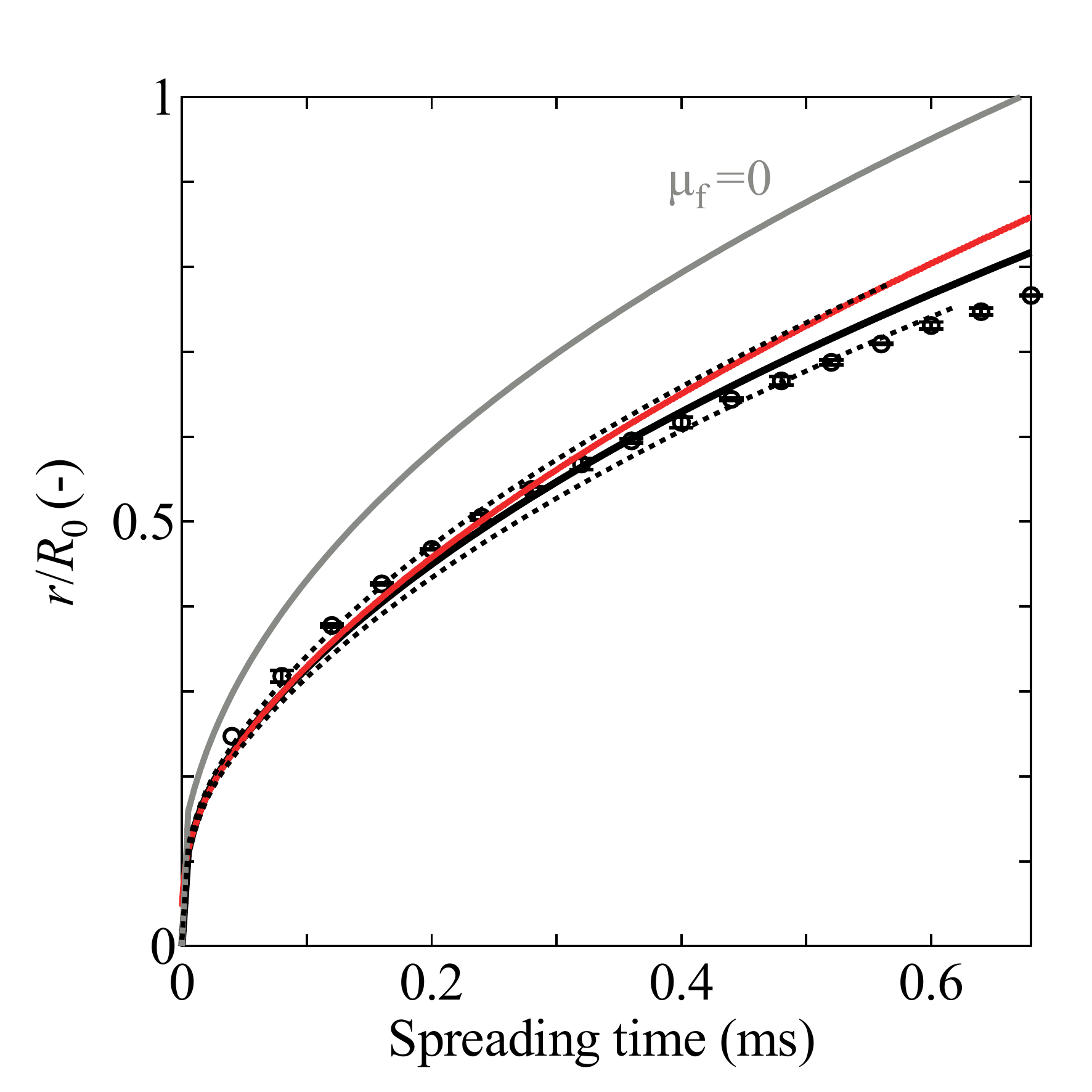}
    \caption{Spreading curves of spontaneous spreading droplets. The numerical curves (black solid lines) are fitted to the experimental spreading curves in order to estimate the friction parameters. The error bars indicate the standard deviations in the experiments.
The dotted lines represent $\pm 20 \%$ deviations from the fitted line friction parameter. The grey line indicate the numerical spreading curve with $\mu_f =0$. The red curve is the case W in TABLE\ref{table:simulations}. }
    \label{fig:A1}
\end{figure}

\bibliography{apssamp}

\end{document}